\documentclass[12pt]{article}
\usepackage{amsmath}
\usepackage{graphicx,psfrag,epsf}
\usepackage{enumerate}
\usepackage{url}
\usepackage{eurosym}
\usepackage{amsmath}
\usepackage{amsfonts}
\usepackage{amssymb}
\usepackage{graphicx}
\usepackage{booktabs}
\usepackage{color}
\usepackage{float}
\usepackage{comment}
\usepackage{rotating}
\usepackage[breaklinks=true,bookmarksopen=true,colorlinks=true,citecolor=blue]{hyperref}
\usepackage{natbib}
\usepackage{setspace}
\usepackage{multirow}
\usepackage{multicol}
\usepackage[utf8]{inputenc}
\usepackage{textcomp}
\usepackage{caption}
\usepackage{subcaption}
\usepackage[table,xcdraw]{xcolor}
\usepackage{adjustbox}

\newcommand{\blind}{0}

\newtheorem{theorem}{Theorem}[section]

\newtheorem{proposition}[theorem]{Proposition}

\begin{document}

\def\spacingset#1{\renewcommand{\baselinestretch}%
{#1}\small\normalsize} \spacingset{1}

\if0\blind
{
  \title{\bf Extending the Scope of Inference About Predictive Ability to
Machine Learning Methods}
  \author{Juan Carlos Escanciano\thanks{
    The author gratefully acknowledges financial support by
MCIN/AEI/10.13039/501100011033  Grants PID2021-127794NB-I00 and CEX2021-001181-M, CAM grants EPUC3M11 (V
PRICIT) and Mad-Econ-Pol-CM H2019/HUM-5891.}\hspace{.2cm}\\
    Department of Economics, Universidad Carlos III de Madrid\\
    and \\
    Ricardo Parra \thanks{The views expressed in this article are exclusively from the author and do not represent the views of neither Banco Santander nor its affiliates.} \\
    Market Risk, Banco Santander CIB}
  \maketitle
} \fi

\if1\blind
{

  \begin{center}
    {\LARGE\bf Extending the Scope of Inference About Predictive Ability to
Machine Learning Methods}
\end{center}
  \medskip
} \fi

\begin{abstract}
{\normalsize The use of machine learning methods for predictive purposes
has increased dramatically over the past two decades, but
uncertainty quantification for predictive comparisons remains elusive. This
paper addresses this gap by extending the classic inference theory
for predictive ability in time series to modern machine
learners, such as the Lasso or Deep Learning. We investigate under which
conditions such extensions are possible. For standard
out-of-sample asymptotic inference to be valid with machine learning, two key properties must hold: (i)
a zero-mean condition for the score of the prediction loss function and
(ii) a \textquotedblleft fast rate\textquotedblright\ of convergence for the
machine learner. Absent any of these conditions, the estimation risk may be unbounded, and inferences invalid and very sensitive to sample splitting. For accurate inferences, we recommend an 80\%-20\% training-test splitting rule.
We illustrate the wide applicability of our results with
three applications: high-dimensional time series regressions with the Lasso, Deep learning for binary outcomes,
and a new out-of-sample test for the Martingale Difference Hypothesis (MDH). The theoretical results are supported by extensive Monte
Carlo simulations and an empirical application evaluating the MDH of
some major exchange rates at daily and higher frequencies. }
\end{abstract}

\vfill

\newpage
\spacingset{1.8}
\section{Introduction}
\label{sec:intro}

Out-of-sample evaluation of predictive models has become a key
feature of modern machine learning methods (see, e.g. %
\citet{james2023introduction}, for a textbook treatment, and %
\citet{mullainathan2017machine}, for an applied econometrics perspective).
It also has a long tradition in time series econometrics, with important
contributions by \citet{diebold1995comparing} and \citet{west1996asymptotic}.
These existing asymptotic justifications for predictive inference provide
a formal framework for quantification of prediction uncertainty. However, these
classical results do not directly carry over to modern machine learning
applications, see, e.g. \citet{masini2023machine} for a review. In this paper, we point out two key properties for valid standard
inference on predictive performance with machine
learning methods: (i) a zero-mean condition for the score of the prediction
loss function; and (ii) a \textquotedblleft fast rate\textquotedblright\ of
convergence for the machine learner. Absent any of these conditions, the estimation risk may be unbounded, and inferences invalid and very sensitive to sample splitting.

The zero-mean condition of the score of the prediction loss
function holds in a number of important examples, including the leading and
canonical example of the Mean Squared Prediction Error (MSPE) loss function
with machine learning estimators for optimal linear projections (e.g. Lasso).%
\footnote{To illustrate, among all papers published at the Journal of Business and
Economic Statistics and Journal of Econometrics in 2008 doing forecast
evaluation, 83\% of them use the MSPE loss function, see Table 1 in %
\citet{gneiting2011making}.} The zero-mean score condition holds when the same loss function
is used for estimation in-sample as for out-of-sample evaluation. Yet, it is common practice to use
different loss functions for estimation and evaluation. Our requirement of
zero-mean score of the prediction loss function at the limit of the
in-sample estimator can be understood as a weak (local) version of the
\textquotedblleft consistent scoring rule condition\textquotedblright\ in %
\citet{gneiting2011making}. This zero-mean condition of the score is also
related to the local robustness property discussed in %
\citet{chernozhukov2022locally}. Our paper goes beyond the debiasing
literature in emphasizing the importance of the fast rates of the machine
learner, even when the zero-mean score property (i.e. local robustness)
holds. Also, we build on the classical literature on predictive ability
tests in time series econometrics, extending its scope to machine learning
methods.

As the main result of the paper, we give general sufficient conditions
under which the estimation risk component of the out-of-sample risk is
asymptotically negligible and standard inference on predictive performance applies,
as in \citet{diebold1995comparing}. In this
standard case, the asymptotic distribution of the out-of-sample risk does
not depend on the limit out-of-sample to in-sample size ratio and the
possibly non-normal asymptotic distribution of the machine learning
estimator. Thus, we provide conditions under which the popular %
\citet{diebold1995comparing} and \cite{west1996asymptotic} forecast
evaluation approaches are simultaneously valid with machine learning
predictions, both leading to the same \textquotedblleft
standard\textquotedblright\ asymptotic inferences. The main practical implications of our
results are that we provide regularity conditions under which (i) commonly
done horse-race predictive comparisons between machine learning methods
and/or classical methods are valid and theoretically justified, see, e.g., \citet{makridakis2018statistical} for an influential study with such comparisons; (ii) these comparisons are less sensitive to sample splitting, and (iii)
estimated asymptotic confidence intervals for the predictive risk, say at
95\%, are valid and have the usual expression of the out-of-sample empirical risk $\pm $
two times the standard error (no correction for estimation risk is
necessary). We recommend reporting confidence intervals, together with point estimates of out-of-sample risk. 

We document theoretically and by simulations the following
points. First, even when the score of the prediction loss function has
zero-mean, the estimation risk part of the out-of-sample empirical risk can
diverge due to the slow rates of the machine learning estimator. Thus, fast
rates for the machine learner are somewhat necessary for standard inference
on predictive performance to be valid. The precise definition of fast rates
is given below in equation (\ref{FastRates}). When both the zero-mean and the fast
rates hold, the estimation risk is asymptotically negligible and standard
inference on the out-of-sample risk applies. We find that the standard
asymptotic theory provides a more accurate approximation with machine
learning methods when the ratio of the out-of-sample period to in-sample
period is smaller than the typical choice of one (a choice of 0.25 works well in our simulations). For mixing
processes, fast rates are attainable for popular machine learning
estimators, such as Lasso, Deep Learning, Boosting, and others, under
suitable regularity conditions, see Section 4 in the Supplementary Material.

The scope of this paper is restricted to machine learning estimators satisfying a fast rate condition and
a class of prediction loss functions whose corresponding risk satisfies a Lipschitz property. This class
includes prominent examples such as the MSPE, the Mean Absolute Deviation (MAD) error, the Huber loss, the ASymmetric MSPE (ASMSPE) loss,
the Log-Cosh loss, the Cross-Entropy loss, and many others. See Section 2.2 below, and Section 2.1 in the Supplementary Material for further discussion on losses.
The machine learner and loss function must also satisfy a stochastic equicontinuity condition, see, e.g., \cite{andrews1994asymptotics} for the same condition. This equicontinuity
condition has been shown to hold in wide generality with machine learners, see, e.g., Theorem 14.20 in \cite{wainwright2019high}.

We illustrate the wide applicability of the general
theory with three applications. First, we consider high-dimensional time
series regressions with the Lasso and the MSPE prediction loss. Second, we study
Deep learning for prediction with a binary outcome using the Cross-Entropy
loss. Finally, we propose a new out-of-sample test for the Martingale
Difference Hypothesis (MDH) with a Ridge estimator and a covariance loss function. We apply it to study the predictability of
some leading exchange rates. We propose a
self-normalized test statistic measuring the sample covariance between the
outcome and the machine learning predictions. In all these applications, we
provide primitive conditions under which the out-of-sample risk converges to a normal
distribution. Extensive Monte Carlo simulations confirm our theoretical
results. An empirical application to exchange rates shows the ability of our
MDH test to detect linear and nonlinear predictability patterns in the data,
complementing traditional Portmanteau tests. Our study contributes to the
extensive and growing literature on the use of machine learning methods for
time series predictability, providing theoretical asymptotic
guarantees for predictive ability with machine learning methods.

Our paper has some important limitations too. One limitation is
that we only consider the case of zero-mean score. We expect that if the
score of the prediction loss function does not have zero-mean the estimation
risk will generally diverge in the high-dimensional setting (even under fast
rates of convergence of the machine learner), though proving so is beyond
the scope of this paper. We report simulations in Section 2.4 of the Supplementary Material supporting this claim.
Another limitation is that we consider stationary mixing data. The analysis of
structural breaks and non-stationarity is also beyond the scope of this paper. Finally,
inferences under slow rates remain unknown in the literature and we are not an exception to this.
We will investigate these interesting extensions in future research.

The rest of the paper is organized as follows. The next section
describes the problem and the main general results. In Section 3, we
consider the application to high-dimensional time series regressions with the Lasso and the MSPE
loss, including Monte Carlo simulations to support our theory. Section 4
reports the application to Deep learning for prediction of binary outcomes
with the Cross-Entropy loss. Section 5 considers the third application to
out-of-sample testing for the MDH based on machine learning predictions.
This section also evaluates the finite sample performance of the new MDH
tests via simulations and an empirical application to exchange rates data.
Finally, Section 6 concludes. The proofs and extensive Monte Carlo sensitivity checks are gathered in a Supplementary
Material.

A word on notation. For a matrix or vector $A,$ $A^{\prime }$
denotes its transpose, $tr(A)$ its trace, $vec(A)$ the
vector obtained by concatenating the column vectors of $A$,
and $\left\Vert A\right\Vert =\sqrt{tr(A^{\prime }A)}$ its Euclidean norm.
If $A$ is positive semidefinite and symmetric, $\lambda _{\max }(A)$ and $%
\lambda _{\min }(A)$ denote its maximum and minimum eigenvalue,
respectively. For a vector $v=(v_{1},...,v_{p})^{\prime },$ $\left\Vert
v\right\Vert _{1}=\sum_{j=1}^{p}\left\vert v_{j}\right\vert $ is the $l_{1}$
norm$\ $and $\left\Vert v\right\Vert _{\infty }=\max_{1\leq j\leq
p}\left\vert v_{j}\right\vert $ is the max norm. Henceforth, $C$ is a
positive constant independent of the sample size and the parameter dimension
$p$, and that may change from expression to expression. When we need more
than one constant in a given expression, we use $C_{1},$ $C_{2},c_{1},$ $c_{2}$,
etc. We will use the acronyms wpa1 and wrt for
\textquotedblleft with probability approaching one\textquotedblright\ and
\textquotedblleft with respect to\textquotedblright , respectively. For two
positive sequences $a_{T}$ and $b_{T},$ we will write $a_{T}\lesssim b_{T}$
when for some $C\in (0,\infty )$ it holds that $\lim \sup a_{T}/b_{T}\leq C$%
. Further, we write $a_{T}\approx b_{T}$ if $a_{T}\lesssim b_{T}\lesssim
a_{T}$. We generalize this notation to random variables and use $a_{T}%
\mathrm{\lesssim }_{\mathbb{P}}b_{T}$ to denote $a_{T}=O_{\mathbb{P}}(b_{T})$.
Henceforth, $R$ and $P=T-R$ are the in-sample and out-of-sample sizes, $T$ is the total sample size, and $%
\hat{E}_{R}$ and $\hat{E}_{P}$ are, respectively, the in-sample and
out-of-sample mean operators%
\begin{equation*}
\hat{E}_{R}\left[ g(Z_{t})\right] :=\frac{1}{R}\sum_{t=1}^{R}g(Z_{t})\qquad
\text{and}\qquad \hat{E}_{P}\left[ g(Z_{t})\right] :=\frac{1}{P}%
\sum_{t=R+1}^{T}g(Z_{t}).
\end{equation*}%

\section{Main General Results}

\label{Main}

We are interested in out-of-sample inference on the moment
\begin{equation*}
E[f_{t}(\theta _{0})],
\end{equation*}%
depending on an unknown $p$-dimensional parameter $\theta _{0}\in \Theta
\subset \mathbb{R}^{p},$ and where the moment function $f_{t}(\theta _{0})$
also depends on observable data at time $t,$ i.e., $f_{t}(\theta
_{0})=f(Z_{t},\theta _{0}).$ The function $f_{t}(\theta _{0})$ is referred
to as the prediction loss function here or the scoring function in %
\citet{gneiting2011making}. The data is a strictly stationary sample $%
\{Z_{t}\}_{t=1}^{T}\ $of sample size $T.$
The parameter $\theta _{0}$ satisfies
\begin{equation}
\theta _{0}\in \arg \min_{\theta \in \Theta }E[h_{t}(\theta )],
\label{ident}
\end{equation}%
for an estimating moment function $h_{t}(\theta ),$ possibly but not
required to be different from $f_{t}(\theta ).$ For a fixed $p,$ this is the
setting considered in \citet{west1996asymptotic}. When $f_{t}(\theta _{0})$
is the loss differential of two forecasts, this is the setting considered in %
\citet{diebold1995comparing}. Our framework can be trivially extended to
multivariate moments, thereby including conditional predictive ability tests
such as those developed in \citet{giacomini2006tests}. We investigate
extensions of this classical theory, permitting a high-dimensional setting
where $p$ is potentially much larger than the sample size $T,$ with $%
p\rightarrow \infty $ as $T\rightarrow \infty ,$ and considering estimators
that are not necessarily asymptotically linear or normal.

\noindent \textbf{Example (MSPE and Lasso):} A leading example of loss function $f_{t}(\theta _{0})$ is the Squared
Prediction Error (SPE) moment function where $f_{t}(\theta _{0})=\left(
Y_{t}-\theta _{0}^{\prime }X_{t}\right) ^{2}\ $and $Z_{t}=(Y_{t},X_{t}^{%
\prime })^{\prime }$. In this example, $X_{t}$ is a $p$-dimensional vector
of predictor variables, possibly including lagged values of the dependent
variable $Y_{t},$ and $\theta _{0}$ is an unknown parameter estimated by a
Lasso estimator with the first $R$ observations, $1\leq R<T,$ as
\begin{equation}
\widehat{\theta }_{R}=\arg \min_{\theta }\hat{E}_{R}\left[ h_{t}(\theta )%
\right] +\lambda _{R}\left\Vert \theta \right\Vert _{1},  \label{Lasso}
\end{equation}%
where $\lambda _{R}$ is a penalization parameter, $\lambda
_{R}\downarrow 0$ as $R\rightarrow \infty ,$ and $h_{t}(\theta )=\left(
Y_{t}-\theta ^{\prime }X_{t}\right) ^{2}$. $\blacksquare $

For simplicity of exposition, we focus on the so-called fixed
forecasting scheme, though our results could potentially be extended to
different forecasting schemes such as the recursive and rolling forecasting
schemes, following the proof of Theorem 1 in \cite{escanciano2010backtesting}%
. Henceforth, we denote by $\widehat{\theta }_{R}$ the
in-sample estimator of $\theta _{0}$ in (\ref{ident}). An
example is the Lasso estimators in (\ref{Lasso}), but other machine learning
estimators are possible.

The expected loss function or risk $E[f_{t}(\theta _{0})]$ is
estimated by the empirical out-of-sample risk $\hat{E}_{P}\left[ f_{t}(%
\widehat{\theta }_{R})\right] $. As in \citet{west1996asymptotic}, we aim to
establish asymptotic distribution theory for
\begin{equation*}
\Delta =\sqrt{P}\left( \hat{E}_{P}\left[ f_{t}(\widehat{\theta }_{R})\right]
-E[f_{t}(\theta _{0})]\right) .
\end{equation*}%
Unlike \citet{west1996asymptotic}, we allow for a high-dimensional setting
where $p$ is potentially much larger than the in-sample estimation period $R$
and/or the out-of-sample period $P.$

A key component in the asymptotic analysis of $\Delta $ is the
Estimation Risk ($ER$) component, defined from the following expansion as%
\begin{equation}
\Delta =\sqrt{P}\left( \hat{E}_{P}\left[ f_{t}(\theta _{0})\right]
-E[f_{t}(\theta _{0})]\right) +\underset{\text{{\small Estimation Risk}}%
\equiv ER}{\underbrace{\sqrt{P}\left( \hat{E}_{P}\left[ f_{t}(\widehat{%
\theta }_{R})-f_{t}(\theta _{0})\right] \right) }}.  \label{ER}
\end{equation}%
A standard Taylor expansion argument yields%
\begin{equation*}
f_{t}(\widehat{\theta }_{R})-f_{t}(\theta _{0})\approx \left[ \dot{f}%
_{t}(\theta _{0})\right] ^{\prime }\left( \widehat{\theta }_{R}-\theta
_{0}\right) ,
\end{equation*}%
where $\dot{f}_{t}(\theta _{0})=\partial f_{t}(\theta _{0})/\partial \theta $
is the score of the loss function, when it exists. Thus, in the classical
fixed $p$ setting, the asymptotic linearity of $\sqrt{P}\left( \widehat{%
\theta }_{R}-\theta _{0}\right) $ yields an asymptotic normal distribution
for the $ER$ under mild moment conditions. The analysis of %
\citet{west1996asymptotic} can be modified to relax the asymptotic linearity
of the estimator, as long as $\sqrt{P}\left( \widehat{\theta }_{R}-\theta
_{0}\right) =O_{\mathbb{P}}(1)$ and $E\left[ \dot{f}_{t}(\theta _{0})\right]
=0,$ since under these conditions%
\begin{equation*}
ER\approx E\left[ \dot{f}_{t}(\theta _{0})\right] ^{\prime }\sqrt{P}\left(
\widehat{\theta }_{R}-\theta _{0}\right) \rightarrow _{\mathbb{P}}0,
\end{equation*}%
where $\rightarrow _{\mathbb{P}}$ denotes convergence in probability.
However, the assumption of $\sqrt{P}-$consistency of $\widehat{\theta }_{R}$
is too strong in the high-dimensional setting. Machine learning estimators
generally have a slower rate of convergence than $P^{-1/2}$. The following subsection shows that even in a stylized example (Gaussian errors and correctly specified linear model), the
$ER$ is no longer asymptotically bounded under the so-called slow rates for
the machine learner.

\subsection{A Stylized Example}

Consider a stylized example of a Gaussian prediction error $%
\varepsilon _{t}=Y_{t}-\theta _{0}^{\prime }X_{t},$ with zero-mean, where $%
\{\varepsilon _{t}\}_{t\in \mathbb{Z}}$ are independent and identically
distributed (iid), independent of $\mathcal{X},$ the $\sigma $-algebra
generated by $\{X_{t}\}_{t=-\infty }^{\infty },$ and with variance $\sigma
^{2}=E[\varepsilon _{t}^{2}].$ Also, $\{X_{t}\}_{t=-\infty }^{\infty }$ are
iid $N(0,\Sigma )$. We consider the SPE loss $f_{t}(\theta
_{0})=\left( Y_{t}-\theta _{0}^{\prime }X_{t}\right) ^{2},$ with a general
machine learning estimator $\widehat{\theta }_{R}$ of $\theta
_{0}.$ Simple algebra shows that, under our assumptions, the
conditional distribution of the $ER$, conditional on $\mathcal{X}$ and $%
\widehat{\theta }_{R},$ is normally distributed, with mean and variance
given, respectively, by%
\begin{equation*}
\left. ER\right\vert \mathcal{X},\widehat{\theta }_{R}\overset{d}{\sim }N(%
\sqrt{P}\hat{r}_{R}^{2},4\sigma ^{2}\hat{r}_{R}^{2}),
\end{equation*}%
where $\overset{d}{\sim }$ means distributed as, $\hat{r}_{R}^{2}=(\widehat{%
\theta }_{R}-\theta _{0})^{\prime }\hat{\Sigma}(\widehat{\theta }_{R}-\theta
_{0})$ and $\hat{\Sigma}=\hat{E}_{P}\left[ X_{t}X_{t}^{\prime }\right] .$
The out-of-sample predictive risk, $\hat{r}_{R}^{2},$ or its expected value $%
r_{R}^{2}=E\left[ \left( X_{t}^{\prime }(\widehat{\theta }_{R}-\theta
_{0})\right) ^{2}\right] ,$ as well as other related quantities such as $%
l_{2}-$rates and in-sample predictive risk, have been extensively
investigated in the machine learning literature, see the references given
below. For example, \citet{greenshtein2004persistence} provide general
conditions for the consistency of Lasso in the iid case in a metric
asymptotically equivalent to $\hat{r}_{R}^{2},$ namely
\begin{equation*}
\tilde{r}_{R}^{2}=(\widehat{\theta }_{R}-\theta _{0})^{\prime }\Sigma (%
\widehat{\theta }_{R}-\theta _{0})\equiv \left\Vert \widehat{\theta }%
_{R}-\theta _{0}\right\Vert _{\Sigma }^{2},
\end{equation*}%
where $\Sigma =E\left[ X_{t}X_{t}^{\prime }\right] $. That is, %
\citet{greenshtein2004persistence} found conditions for $\tilde{r}%
_{R}^{2}\rightarrow 0$ as $R\rightarrow \infty$ (in which case the estimator is called persistent).

{\normalsize Next, it is natural to study the rates at which $r_{R}^{2},$ }$%
\tilde{r}_{R}^{2}$ {\normalsize or} {\normalsize $\hat{r}_{R}^{2}$ go to
zero with }$R.${\normalsize \ Under general conditions on predictors only
the so-called slow rates are possible, see, e.g., \citet{rigollet2011exponential} and %
\citet{dalalyan2017prediction}, where
\begin{equation}
r_{R}^{2}\approx \frac{\left\Vert \theta _{0}\right\Vert _{1}\sqrt{\log p}}{%
\sqrt{R}}.  \label{A2}
\end{equation}%
Therefore, in the common practical case where the relative out-of-sample
size to in-sample size is positive, i.e. $\lim_{T\rightarrow \infty }P/R>0,$
one has $\sqrt{P}r_{R}^{2}\rightarrow \infty .$ Our first observation is
that under general conditions on the sample splitting ($R$ is not
necessarily large relative to $P$) and the parameters }$(\theta _{0},\Sigma
,\sigma ),${\normalsize \ the $ER$ diverges. Define the }$l_{1}$
{\normalsize ball with radius }$r>0,${\normalsize \ }$B_{1}(r)=\left\{ v\in
\mathbb{R}^{p}:{\normalsize \left\Vert v\right\Vert _{1}\leq r}\right\} $%
{\normalsize \ and emphasize the dependence of }$\hat{r}_{R}^{2}$%
{\normalsize \ on }$\widehat{\theta }_{R}$ {\normalsize and }$\theta _{0},$%
{\normalsize \ }$\hat{r}_{R}^{2}(\widehat{\theta }_{R}{\normalsize ,}\theta
_{0}).$

\begin{proposition}
{\normalsize \label{Prop1}In this example, for a persistent procedure (i.e. $%
\hat{r}_{R}\rightarrow _{\mathbb{P}}0$), if $\sqrt{P}\hat{r}%
_{R}^{2}\rightarrow _{\mathbb{P}}\infty ,$ then $ER\rightarrow _{\mathbb{P}%
}+\infty .$ Moreover,}%
\begin{equation}
\inf_{\widehat{\theta }_{R}}\sup_{\theta _{0}\in B_{1}(r)}\sqrt{P}%
{\normalsize \hat{r}}_{R}^{2}{\normalsize (}\widehat{\theta }_{R}%
{\normalsize ,\theta }_{0}{\normalsize )\gtrsim }_{\mathbb{P}}{\normalsize %
\lambda }_{\min }{\normalsize (\Sigma )r\sigma }\sqrt{\frac{P\log p}{R}.}
\label{minimax}
\end{equation}%
{\normalsize Thus, if the right hand side diverges, then $ER\rightarrow _{%
\mathbb{P}}+\infty .$}
\end{proposition}

{\normalsize \noindent This simple observation shows that even in a
favorable situation where the model is correctly specified and the score of
the loss function has zero-mean, the $ER$ will diverge in general. This
result has important practical implications because often researchers choose
$P$ of a similar magnitude to $R$ (two-fold validation with equal sample
sizes), which implies the condition $\sqrt{P}r_{R}^{2}\rightarrow \infty \ $%
holds in the presence of slow (minimax) rates for machine learners (i.e. when the
machine learner satisfies (\ref{minimax}) with equality). The second part of
Proposition \ref{Prop1} gives a lower bound that holds for any machine
learning estimator (the $\inf$ is over all possible estimators). This result follows from \citet{raskutti2011minimax}.
Therefore, the problem we point out is not specific to the Lasso but it is rather related to the fundamental or
information-theoretic limitations of the statistical problem at hand and
apply to general machine learners.}
%

Therefore, some additional assumptions are necessary to achieve
faster rates for machine learners. These additional conditions have been
well investigated in the literature, see the so-called \textquotedblleft
restricted eigenvalue conditions\textquotedblright\ for fast rates of
convergence for Lasso estimators in, e.g., \citet{bickel2009simultaneous}
and \citet{van2009conditions}, or smoothness and compositional adaptive rates for Deep learning in
\cite{schmidt2020nonparametric}, among others.

\subsection{The case of zero-mean score and fast rates}

{\normalsize In this section we consider a relatively general setting that
leads to valid asymptotic predictive inference in a high-dimensional
framework. We introduce high-level assumptions here and provide low-level,
primitive, conditions in Sections 3, 4 and 5 for important examples. Let }$%
F_{Z}$ {\normalsize denote the cumulative distribution function of the
random vector $Z_{t}=(Y_{t},X_{t})$}$.$ {\normalsize Define the mean
operator }$E_{Z}\left[ f_{t}(\widehat{\theta }_{R} )\right] =\int f(\widehat{\theta }_{R} ,{\normalsize z}%
)dF_{Z}(z)$ (note this is random). {\normalsize Henceforth, we consider loss functions of the form
}$f_{t}(\theta )=\ell(Y_{t},m(\theta ,X_{t})),${\normalsize \ for a loss
function }$\ell${\normalsize \ and a predictive model }$m(\theta ,X_{t})$
{\normalsize satisfying our conditions below.}

{\normalsize \noindent\textbf{Assumption A1: }$R,P\rightarrow\infty$ as $%
T\rightarrow \infty,$ and $\lim_{T\rightarrow\infty}P/R=\pi,$ $%
0\leq\pi<\infty$. }

{\normalsize \noindent \textbf{Assumption A2}:\textbf{\ }(Stochastic
Equicontinuity, SE) The estimator and loss function satisfy the SE condition}%
\begin{equation*}
\left\vert \sqrt{P}\left( \hat{E}_{P}\left[ f_{t}(\widehat{\theta }%
_{R})-f_{t}(\theta _{0})\right] -E_{Z}\left[ f_{t}(\widehat{\theta }%
_{R})-f_{t}(\theta _{0})\right] \right) \right\vert \rightarrow _{\mathbb{P}%
}0.
\end{equation*}%

{\normalsize \noindent \textbf{Assumption A3}:\textbf{\ (}Lipschitz risk and
zero-mean score). The loss function satisfies}%
\begin{equation}
C_{1}E\left[ \left( m(\theta ,X_{t})-m(\theta _{0},X_{t})\right) ^{2}%
\right] \leq E\left[ f_{t}(\theta )-f_{t}(\theta _{0})\right] \leq
C_{2}E\left[ \left( m(\theta ,X_{t})-m(\theta _{0},X_{t})\right) ^{2}%
\right] ,  \label{Lipschitz}
\end{equation}%
{\normalsize for all }$\theta ,\theta _{0}${\normalsize $\in \Theta $ and
some positive constants $C_{1}$ and $C_{2}.$ Moreover, }%
\begin{equation}
\frac{\partial E\left[ f_{t}(\theta _{0})\right] }{\partial \theta }=0.
\label{zero-mean}
\end{equation}%
{\normalsize \noindent \textbf{Assumption A4}:\textbf{\ (}Fast rates). The
estimator }$\widehat{\theta }_{R}$ {\normalsize has a fast rate in the sense
that}%
\begin{equation}
\sqrt{P}E_{Z}\left[ \left( m(\widehat{\theta }_{R},X_{t})-m(\theta
_{0},X_{t})\right) ^{2}\right] \rightarrow _{\mathbb{P}}0. \label{FastRates}
\end{equation}

{\normalsize \noindent Assumptions A1-A4 are standard in the literature,
see, e.g., \cite{andrews1994asymptotics}. We aim for asymptotic results that do not rely on $\pi=0$. We verify Assumptions A2, A3 and
A4 for several examples in Sections 3, 4 and 5. These assumptions allow for
a wide class of loss functions and estimators. A key assumption is the
zero-mean score (\ref{zero-mean}), which often follows from (\ref{Lipschitz}). We provide the main
result of this section. }

\begin{proposition}
{\normalsize \label{Prop2}Under Assumptions A1-A4, $ER\rightarrow _{\mathbb{P%
}}0.$ }
\end{proposition}

{\normalsize \noindent \textbf{Example (MSPE and Lasso, cont.):} This
example corresponds to }$\ell(y,m)=(y-m)^{2}$ {\normalsize and }$m(\theta
,X_{t})=\theta ^{\prime }X_{t}.$ {\normalsize We give in Section \ref%
{SectionLasso} primitive conditions for the Lasso estimator to satisfy
Assumptions A2 and A4, and the zero-mean condition corresponding to this
loss. Assumption A3 holds with }$C_{1}=C_{2}=1.$ ${\normalsize \blacksquare }
$

Table \ref{Table0} provides examples of loss functions
satisfying our Assumption A3. For simplicity, we consider for the moment the case where
 $\ell(Y_{t},m(\theta ,X_{t}))$ is a function of $\varepsilon _{t}=Y_{t}-m(\theta ,X_{t})$. However, we
note that our results are also applicable more generally. In all these
cases, %
\begin{equation*}
\frac{\partial E\left[ f_{t}(\theta _{0})\right] }{\partial \theta }=E%
\left[ \psi (\varepsilon _{t})\dot{m}(\theta _{0},X_{t})\right] ,
\end{equation*}%
for a suitable function $\psi (\varepsilon _{t})$ given in Table \ref{Table0}, and $\dot{m}(\theta _{0},X_{t})=\partial m(\theta
_{0},X_{t})/\partial \theta .$ Thus, the zero-mean score
condition means that $E\left[ \psi (\varepsilon _{t})\dot{m}(\theta
_{0},X_{t})\right] =0.$ The last column of the table verifies
the zero-mean condition under the assumption that the conditional
distribution of $\varepsilon _{t}$ given $X_{t}$ is symmetric around zero. See Section 2.1 in the Supplementary Material for further discussion. Henceforth, $\bold{1}(A)$ denotes the indicator of the event $A,$
equals one if $A$ holds, $cosh$ and $tanh$ denote the hyperbolic cosine and tangent functions, respectively, and sign($\cdot$) denotes the sign function.

\begin{equation*}
\fbox{TABLE 1 ABOUT HERE}
\end{equation*}%

For binary classification, a popular loss function is the Cross-Entropy or
log-likelihood loss, corresponding to $\ell(y,m)=-ym+\log (1+e^{m}),$%
{\normalsize \ where }$y\in \{0,1\}$ {\normalsize is binary. This example
has a mean score }$E\left[ \varepsilon _{t}\dot{m}(\theta
_{0},X_{t})\right] ,$ {\normalsize where now }$\varepsilon _{t}=Y_{t}-\Lambda
(m(\theta _{0},X_{t})),$ {\normalsize and }$\Lambda (u)=\exp (u)/1+\exp (u)$
{\normalsize is the Logit distribution.} {\normalsize This score has zero-mean
when }$\theta _{0}$ is {\normalsize estimated by the conditional likelihood
estimator or penalized versions of it.} Another commonly used loss is the
covariance loss function $\ell(y,m)=ym$, which has zero-mean score when $E\left[ Y_{t}\dot{m}(\theta
_{0},X_{t})\right] =0.$ This holds when $Y_{t}$ is a martingale difference sequence, see Section 5.

{\normalsize All these examples satisfy the Lipschitz condition in (\ref%
{Lipschitz}) under well-known conditions. For example, for the MAD this
requires that the conditional density of }$Y_{t}${\normalsize \ given }$%
X_{t} ${\normalsize \ is bounded and bounded away from zero by constants }$%
C_{1}${\normalsize \ and }$C_{2},${\normalsize \ respectively. The zero-mean
condition is satisfied when the same loss function is used for estimation
and for forecast evaluation, but it also holds generally when the error
distribution is (conditionally) symmetric. When the error distribution is
not symmetric and the machine learning estimator estimates the optimal
linear predictor, the zero-mean score condition may not be satisfied, as we
illustrate in Section 2 in the Supplementary Material with further discussion and simulations. The results for standard inference to
be valid with machine learning do not apply in this case, which is beyond
the scope of this paper.}

{\normalsize In the next sections, we verify Assumptions A2 and A4 under
primitive conditions for several examples, starting with high-dimensional
time series regressions with the Lasso and the MSPE loss function. }

\section{High-dimensional time series Lasso regressions}

\label{SectionLasso}

\subsection{Theoretical developments}

{\normalsize \noindent In what follows we establish primitive conditions for
Assumptions A2 and A4 to hold for the Lasso and the MSPE, i.e. $f_{t}(\theta
)=h_{t}(\theta )=\left( Y_{t}-\theta ^{\prime }X_{t}\right) ^{2}\ $and }$%
\widehat{\theta }_{R}${\normalsize \ as in (\ref{Lasso})},{\normalsize \
building on recent work by \citet{wong2020lasso}. We introduce notation and
assumptions for the asymptotic results. Define the $\beta $-mixing
coefficients for a time series process $\{Z_{t}\}$ as
\begin{equation*}
\beta (n)=\frac{1}{2}\sup \sum_{i=1}^{I}\sum_{j=1}^{J}\left\vert P(A_{i}\cap
B_{j})-P(A_{i})P(B_{j})\right\vert ,\text{ }n\geq 1,
\end{equation*}%
where the supremum is over all pairs of partitions $\{A_{1},...,A_{I}\}$ and
$\{B_{1},...,B_{J}\}$ of the sample space such that $B_{j}\in \mathcal{F}%
_{T} $ and $A_{i}\in \mathcal{P}_{T+n},$ and the $\sigma $-fields $\mathcal{F%
}_{T} $ and $\mathcal{P}_{T}$ are defined as $\mathcal{F}_{T}:=\sigma
(Z_{t},t\leq T)$ and $\mathcal{P}_{T}:=\sigma (Z_{t},t\geq T),$ respectively$%
.$ Define the sub-Weibull($\gamma $) norm of a $p$-dimensional random vector
$S_{t},$ for $\gamma >0,$ as%
\begin{equation*}
\left\Vert S_{t}\right\Vert _{\psi _{\gamma }}=\sup_{\delta \in \mathbb{S}%
^{p-1}}\sup_{q\geq 1}\left( E\left[ \left\vert S_{t}^{\prime }\delta
\right\vert ^{q}\right] \right) ^{1/q}q^{-1/\gamma },
\end{equation*}%
where $\mathbb{S}^{p-1}$ is the unit sphere in $\mathbb{R}^{p}.$ }

{\normalsize \noindent \textbf{Assumption A5: }$\{Y_{t},X_{t}^{\prime
}\}_{t\in \mathbb{Z}}$ is strictly stationary and $\beta$-mixing process with
mixing coefficients satisfying for some constant $\gamma _{1}>0,$ $\beta (n)\leq 2\exp \left( -Cn^{\gamma _{1}}\right), n\in \mathbb{N},$
for a constant $\gamma _{2}>0,$ $\left\Vert S_{t}\right\Vert _{\psi
_{\gamma _{2}}}<C$, for $S_{t}=Y_{t}$ and $S_{t}=X_{t}.$ }

{\normalsize \noindent Assumption A5 is from \citet{wong2020lasso} and it is
extensively discussed there. We also need the following assumption. We say that $\theta _{0}$ is $%
s-$sparse if $\theta _{0}$ has at most }$s${\normalsize \ non-zero components}. Recall $%
\Sigma =E\left[ X_{t}X_{t}^{\prime }\right]$ and $\lambda
_{R}$ is the Lasso penalization parameter.

{\normalsize \noindent \textbf{Assumption A6}:\textbf{\ }(i) $\lambda
_{R}\approx \sqrt{\frac{\log p}{R}};\ $(ii) for each }$p,$ $0<\lambda _{\min
}(\Sigma )\leq ${\normalsize \ $\lambda _{\max }(\Sigma )\leq C<\infty $; (iii) $%
\theta _{0}$ is $s-$sparse with
\begin{equation}
\frac{s}{\lambda_{\min
}(\Sigma ) }\frac{\log p}{\sqrt{R}}=o(1).  \label{FR}
\end{equation}%
}

{\normalsize We also need some further notation. For $a_{t}=f_{t}(\theta
_{0})-E[f_{t}(\theta _{0})],$ define $\Gamma _{a}(j)=E[a_{t}a_{t-j}]$ and $%
\Omega _{a}=\sum_{j=-\infty }^{\infty }\Gamma _{a}(j)$. Assumption A5
implies that the previous long-run variance exists. With this notation in
place, we can derive the following result. }

\begin{theorem}
{\normalsize \label{Thm1A}\textit{Under Assumptions A1, A5 and A6, }$%
ER\rightarrow _{\mathbb{P}}0$ and\textit{\ }
\begin{equation*}
\frac{1}{\sqrt{P}}\sum_{t=R+1}^{T}(f_{t}(\widehat{\theta }%
_{R})-E[f_{t}(\theta _{0})])\overset{d}{\longrightarrow }N(0,\Omega _{a}),
\end{equation*}%
for $f_{t}(\theta )=\left( Y_{t}-\theta ^{\prime }X_{t}\right) ^{2}$ and the Lasso estimator $%
\widehat{\theta }_{R}${\normalsize \ as in (\ref{Lasso})}}.
\end{theorem}

{\normalsize \noindent It is important to note that the limiting
distribution does not depend on the non-Gaussian limiting
distribution of the Lasso estimator in (\ref{Lasso}) or the forecasting scheme
(e.g. on $\lim_{T\rightarrow \infty }P/R)$. The asymptotic distribution is
the same as the one in \citet{diebold1995comparing}, since the $ER$ is
asymptotically negligible. }

\subsection{Monte Carlo Experiments}

{\normalsize Several Monte Carlo experiments are conducted in order to check
the finite sample performance of our asymptotic theory. First, we will
consider a Data Generating Process (DGP) in which sparsity is of the same
order as the root of the number of parameters, resulting in slow rates of
the machine learner (here Lasso) and a divergence of the $ER$; we refer to
this case as the Decreasing Sparsity case. Then, we will examine a DGP in
which multicollinearity is introduced, leading to a similar divergence of
the $ER$. Finally, we will present a DGP in which the Lasso estimator
attains fast rates and the $ER$ converges to 0. For completeness, we provide
an analysis of the eigenvalues of the covariance matrix $\Sigma =E\left[
X_{t}X_{t}^{\prime }\right] $ in Section 3 of the Supplementary Material.}

\subsubsection{Decreasing Sparsity}

{\normalsize Consider a simple linear model in matrix form as: $Y=X\theta
_{0}+\varepsilon$, where each column of the $R\times p$ matrix $X$ contains
draws from i.i.d. standard Normal random variables, $\varepsilon $ is an $%
R\times 1$ vector of independent $N(0,1),$ and $\theta _{0}$ is a vector
with length equal to the total sample size ($p=T$) such that its first $s$
elements alternate between 1 and -1 while the rest are $0$. The parameters
are estimated by Lasso with a regularization parameter $\lambda _{R}=\sqrt{%
\frac{log(p)}{R}}$. For each Monte Carlo replication (500 for each DGP), we
compute the $ER$ and $\Delta$ in (\ref{ER}).

A key variable for the convergence of the Lasso and its $ER$ is the sparsity
of the data. We set the sparsity index to $s=\lceil \sqrt{p}-27\rceil $,
where $\lceil x\rceil $ is the ceiling function, the smallest integer
greater than or equal to $x,$ corresponding to $s=5$, $s=18$, and $s=50$ for
$T=1000$, $T=2000$ and $T=6000,$ respectively. We refer to this case as the
decreasing sparsity case because $s$ increases (and $s/p$ decreases) with
the number of parameters $p$. This level of sparsity is weak enough to make
the $ER$ and $\Delta $ to diverge with $T$, as one can see from Figure \ref%
{Figure1} and Figure \ref{Figure2}. As the sample size increases, the bias
of both $\Delta $ and $ER$ increases. As a consequence, standard asymptotic
inference as in \citet{west1996asymptotic} will be invalid. The left plot of
Figure \ref{Figure1} considers the common case in applications where $R=P\ $(%
$\pi =1$). We observe large biases in the $ER\ $for this case, consistent
with large values of $\sqrt{P}\hat{r}_{R}^{2}$ in our stylized example of
Section 2.1.
\begin{equation*}
\fbox{FIGURES 1 AND 2 ABOUT HERE}
\end{equation*}
}

\subsubsection{Multicollinearity}

{\normalsize We keep everything like in the previous DGP, except for the
explanatory variables being generated and the sparsity index. Now, the first
variable will be drawn from a standard normal and the rest will be
constructed adding a $N(0,0.1)$ to the first variable. This will create
multicollinearity, which can break the restricted eigenvalue conditions and
prevent the Lasso from attaining fast rates. The sparsity index will be
fixed at 15. Figure \ref{Figure3} and Figure \ref{Figure4} show plots for $%
\Delta $ and the Estimation Risk in this case. Standard inference is again
invalid due to the large biases in the $ER.$
\begin{equation*}
\fbox{FIGURES 3 AND 4 ABOUT HERE}
\end{equation*}
}

\subsubsection{Fast Rates}

{\normalsize We consider now the same DGP as in the decreasing sparsity case
but now we fix the sparsity index at 5. In this framework, Lasso attains
fast rates so the $ER$ converges to 0 in probability and therefore $\Delta $
converges to a zero-mean normal distribution, see Figures \ref{Figure5} and \ref{Figure6}.
\begin{equation*}
\fbox{FIGURES 5 AND 6 ABOUT HERE}
\end{equation*}%
For completeness, we also report in Table \ref{Table 1} the coverage
probabilities for standard asymptotic confidence intervals at different
nominal levels for the three DGPs. As explained by the theory, when the
ratio of out-of-sample to in-sample size, $\pi ,$ is small, the bias coming
from the estimation risk is less relevant. This agrees with the fact that,
holding everything else constant, we get better coverage for the confidence
intervals with $\pi =0.25$ than with $\pi =1$. Only for the case of fast
rates the coverage is close to the nominal level and less sensitive to $\pi$ (for further results on sensitivity to $\pi$ see Section 2 in the Supplementary Material).
\begin{equation*}
\fbox{TABLE 2 ABOUT HERE}
\end{equation*}%
}

\section{Application to Deep Learning for binary
prediction}

{\normalsize As a second application of our results, we consider
penalized Deep learning for binary prediction. As the loss function,
we use the Cross-Entropy or log-likelihood loss }$\ell(y,m)=-ym+\log (1+e^{m}),${\normalsize \
where }$y\in \{0,1\}$ {\normalsize is binary, and }$%
m(\theta ,x)$ {\normalsize is a Deep Neural Network (DNN) model with an architecture }%
$(L,w),${\normalsize \ where }$L${\normalsize \ stands for the number of
hidden layers or depth, and }$w=(w_{0},w_{1},...,w_{L+1})${\normalsize \
denotes the width vector. The DNN can be expressed as}%
\begin{equation}
m(\theta ,x)=A_{L+1}\circ \sigma _{L}\circ A_{L}\circ \sigma _{L-1}\circ
\cdots \circ \sigma _{1}\circ A_{1}(x),  \label{DNN}
\end{equation}%
{\normalsize where }$A_{j}:\mathbb{R}^{w_{j-1}}\rightarrow \mathbb{R}^{w_{j}}
${\normalsize \ is a linear affine map, defined by }$A_{j}(x)=W_{j}x+b_{j},$%
{\normalsize \ for a given }$w_{j-1}\times w_{j}${\normalsize \ weight
matrix }$W_{j}${\normalsize \ and bias vector }$b_{j}\in \mathbb{R}^{w_{j}},$%
{\normalsize \ }$\sigma _{j}:\mathbb{R}^{w_{j}}\rightarrow \mathbb{R}^{w_{j}}
${\normalsize \ is a nonlinear element-wise activation function, which we
take for concreteness as the ReLU (rectilinear unit) function}$,$%
{\normalsize \ i.e. }$\sigma _{j}(z)=(\sigma (z_{1}),...,\sigma
(z_{w_{j}}))^{\prime }${\normalsize \ with }$\sigma (z_{1})=\max (0,z_{1})$%
{\normalsize \ and }$z=(z_{1},...,z_{w_{j}})^{\prime }.${\normalsize \ The
parameter }$\theta ${\normalsize \ gathers all the parameters of the model}%
\begin{equation*}
\theta =\left( vec(W_{1})^{\prime },b_{1}^{\prime },...,vec(W_{L+1})^{\prime
},b_{L+1}^{\prime }\right) ^{\prime }.
\end{equation*}%
{\normalsize With }$m(\theta ,x)${\normalsize \ as in (\ref{DNN}), define
the set of predictive DNN models}%
\begin{equation*}
\mathcal{M}_{L,N,B,F}=\left\{ m(\theta ,x):\max_{1\leq j\leq L}\left\vert
w_{j}\right\vert \leq N,{\normalsize \left\Vert \theta \right\Vert _{\infty }%
}\leq B,\sup_{x}\left\vert m(\theta ,x)\right\vert \leq F\right\} .
\end{equation*}%
{\normalsize With this notation in place, we define the penalized DNN estimator }%
\begin{equation}
\widehat{\theta }_{R}=\arg \min_{\theta :m(\theta ,)\in \mathcal{M}%
_{L,N,B,F}}\hat{E}_{R}\left[ \ell(Y_{t},m(\theta ,X_{t}))\right] +\lambda
_{R}\left\Vert \theta \right\Vert _{clip,\tau _{R}},  \label{PDNN}
\end{equation}%
{\normalsize where }$\lambda _{R}${\normalsize \ is a penalization parameter
such that }$\lambda _{R}\downarrow 0${\normalsize \ as }$R\rightarrow \infty
,${\normalsize \ }$\ell(y,m)=-ym+\log (1+e^{m})${\normalsize , }$\left\Vert
\theta \right\Vert _{clip,\tau }=\sum_{j=1}^{p}${\normalsize $\min \left(
\left\vert \theta _{j}\right\vert /\tau ,1\right) $ is the clipped }$L_{1}$%
{\normalsize \ norm, $\theta =(\theta _{1},...,\theta _{p})^{\prime },$ and }%
$\tau _{R}>0$ {\normalsize is a clipping threshold. Here the tuning
parameters }$(L,N,B,F,\lambda ,\tau )${\normalsize \ are all allowed to
depend on the in-sample size }$R${\normalsize , with rates specified in the
following results.}

{\normalsize In what follows, we establish primitive conditions for
Assumptions A2 and A4 to hold for this example, building on recent work by
\cite{kengne2024deep}. Other applications of our results for Deep
learning and mixing sequences, such as, for example, for MSPE loss and
additive regression with diverging number of components,
can be obtained from the work of \cite{deb2024trade}. }

{\normalsize We introduce notation and assumptions for the asymptotic
results. Define the }${\normalsize \alpha }${\normalsize -mixing
coefficients for a time series process $\{Z_{t}\}$ as
\begin{equation*}
\alpha (n)=\sup \left\vert P(A\cap B)-P(A)P(B)\right\vert ,\text{ }n\geq 1,
\end{equation*}%
where the supremum is over all $B\in \mathcal{F}_{T}$ and $A\in \mathcal{P}%
_{T+n}$. Define the prediction error
term}%
\begin{equation*}
\varepsilon _{t}=Y_{t}-\frac{\exp (m(\theta _{0},X_{t}))}{1+\exp (m(\theta
_{0},X_{t}))},
\end{equation*}%
{\normalsize and let }$\mathcal{C}^{s}(\mathcal{X},\mathcal{K})$
{\normalsize denote the class of }$s-${\normalsize H\"{o}lder functions on }$%
\mathcal{X}\in\mathbb{R}^d,$ {\normalsize the support of }$X_{t},$ {\normalsize with radius
}$\mathcal{K}>0;$ {\normalsize see the definition before the proof of Theorem 4.1 in the Supplementary Material.} The following assumptions are considered in \cite{kengne2024deep}.

{\normalsize \noindent \textbf{Assumption A7: }$\{Y_{t},X_{t}^{\prime
}\}_{t\in \mathbb{Z}}$ is strictly stationary and $\alpha$-mixing process with
mixing coefficients satisfying for some constant $\gamma >0,$ $\alpha (n)\leq {\normalsize C_{1}}\exp \left( -C_{2}n^{\gamma }\right) ,%
\text{ }n\in \mathbb{N},$
for constants $C_{1},{\normalsize C_{2}}>0.$ The support }$\mathcal{X}$%
{\normalsize \ is a compact set of }$\mathbb{R}^{d}.$ {\normalsize Moreover,
}$E[\left. \varepsilon _{t}\right\vert \mathcal{F}_{t-1}]=0${\normalsize \
a.s. and }$m(\theta _{0},\cdot )\in \mathcal{C}^{s}(\mathcal{X},\mathcal{K}).
$

{\normalsize \noindent \textbf{Assumption A8}:\textbf{\ }(i) $\lambda
_{R}\approx R^{-\frac{\gamma }{\gamma +1}}\log (R)$ and }$\tau _{R}\lesssim
(L_{R}+1)^{-1}\left[ (N_{R}+1)B_{R}\right] ^{-L_{R}-1}R^{-\frac{\gamma }{%
\gamma +1}}${\normalsize $;\ $(ii) }$L_{R}\approx \log (R),$ $N_{R}\lesssim
R^{\frac{\gamma c_{1}}{\gamma +1}},$ $1\lesssim B_{R}\lesssim R^{\frac{%
\gamma c_{2}}{\gamma +1}},$ {\normalsize for }$c_{1},c_{2}>0${\normalsize ;
(iii) the following rate condition holds for a }$v>3,${\normalsize \
\begin{equation}
\sqrt{\frac{P}{R}}R^{-\frac{\gamma s}{(\gamma +1)(s+d)}+\frac{1}{2}}\left[
\log R\right] ^{v}=o(1).  \label{FRDNN}
\end{equation}%
}

\begin{theorem}
{\normalsize \label{Thm1B}\textit{Under Assumptions A1, A7 and A8, }$%
ER\rightarrow _{\mathbb{P}}0$ and\textit{\ }
\begin{equation*}
\frac{1}{\sqrt{P}}\sum_{t=R+1}^{T}(f_{t}(\widehat{\theta }%
_{R})-E[f_{t}(\theta _{0})])\overset{d}{\longrightarrow }N(0,\Omega _{a}),
\end{equation*}%
for $f_{t}(\theta )=-Y_{t}m(\theta ,X_{t})+\log (1+e^{m(\theta ,X_{t})})$ and $\widehat{\theta }_{R}$%
{\normalsize \ as in (\ref{PDNN})}}
\end{theorem}

{\normalsize \noindent Fast rates for the DNN estimator hold under the key rate condition}%
\begin{equation}
\frac{\gamma }{\gamma +1}>\frac{s+d}{2s},  \label{R1}
\end{equation}%
{\normalsize which shows a trade-off between dependence and complexity of
the class (as measured by the smoothness index }$s${\normalsize \ and the
dimension of covariates }$d).$ {\normalsize For example, in the iid case, }$%
\gamma =\infty $,{\normalsize \ the level of smoothness }$s${\normalsize %
\ must be larger than }$d.$ An {\normalsize implication of (\ref{R1}) is }$%
\gamma >1${\normalsize \ and }$s>(\gamma +1)d/(\gamma -1),${\normalsize \ so
the level of smoothness must be sufficiently large for the fast rate of
convergence to hold. Assumptions like these are standard in the literature
(see e.g. \cite{andrews1994asymptotics}).}

\section{Testing the Martingale Difference Hypothesis}

\subsection{The new test and its asymptotic distribution}

{\normalsize Another illustration of the wide applicability of the previous
results is a new out-of-sample test for the Martingale Difference
Hypothesis (MDH), as in \citet{clark2006using} but in a high-dimensional
setting. The MDH is one the most prominent hypothesis in time series
econometrics, with a long history; see \citet{escanciano2009b} for a review.
Despite the long history and developments, we are not aware of any
theoretical justification of the use of machine learning predictive methods
for its out-of-sample evaluation. To the best of our knowledge, ours is the
first formal out-of-sample MDH test allowing for machine learners in a time
series setting. }

{\normalsize This application considers the loss function suggested in %
\citet{clark2006using},
\begin{equation*}
f_{t}(\theta _{0})=Y_{t}^{2}-\left( Y_{t}-\theta _{0}^{\prime }X_{t}\right)
^{2}+\left( \theta _{0}^{\prime }X_{t}\right) ^{2}\equiv 2Y_{t}X_{t}^{\prime
}\theta _{0}
\end{equation*}%
Let $\mathcal{F}_{t}$ denote the $\sigma $-field generated by $Y_{t},$ and
possibly other variables $X_{t},$ $\mathcal{F}_{t}:=\sigma (Z_{s},s\leq t),$
$Z_{t}=(Y_{t},X_{t}).$ The vector $X_{t}$ includes an intercept, $X_{t}\in
\mathcal{F}_{t-1},$ and is possibly high-dimensional. For example, $%
X_{t}=(1,Y_{t-1},...,Y_{t-p})^{\prime }$ for a large $p\geq 1.$ Like in %
\citet{clark2006using}, we aim to test the null hypothesis that $Y_{t}$ is a
martingale difference sequence wrt $\mathcal{F}_{t-1},$ i.e.%
\begin{equation}
E[\left. Y_{t}\right\vert \mathcal{F}_{t-1}]=0\text{ a.s.}  \label{H0}
\end{equation}%
against the alternative that $E[\left. Y_{t}\right\vert \mathcal{F}_{t-1}]\neq 0.$
In the fixed dimensional case, \citet{clark2006using} used an asymptotic
theory where $R$ and $\widehat{\theta }$ are fixed, while $P\rightarrow
\infty .$ In contrast, we allow for both $R$ and $P\rightarrow \infty ,$ $%
\widehat{\theta }$ is random and possibly estimated by high-dimensional
methods, such as the Ridge estimator defined below. We will employ the fact
that the sample version of the loss function $f_{t}(\theta
_{0})=2Y_{t}X_{t}^{\prime }\theta _{0},$ suitably standardized, converges by
the martingale central limit theorem (CLT) to a standard normal under the
null hypothesis of the MDH. In this application,
\begin{equation*}
\Delta =\sqrt{P}\left( \hat{E}_{P}\left[ 2Y_{t}X_{t}^{\prime }\widehat{%
\theta }\right] -E[2Y_{t}X_{t}^{\prime }\theta _{0}]\right) ,
\end{equation*}%
where $\widehat{\theta }$ is an estimator based on in-sample observations,
such as, for example, the Ridge estimator
\begin{equation}
\widehat{\theta }=\arg \min_{\theta }\hat{E}_{R}\left[ \left( Y_{t}-\theta
^{\prime }X_{t}\right) ^{2}\right] +\lambda _{R}\left\Vert \theta
\right\Vert ^{2}.  \label{Ridge}
\end{equation}%
One may argue that a Ridge estimator is better motivated than a Lasso
estimator for testing the MDH with time series. It is well-known that Ridge
performs better than Lasso in dense models where some of the coefficients,
here related to the moments $E[Y_{t}X_{t}]$, are expected to be small but
not exactly zero. Nevertheless, we allow for a generic estimator $\widehat{%
\theta },$ not necessarily Ridge, with the requirement that the estimator
does not have a probability mass at zero under the null of the MDH. This is
necessary to avoid dividing by zero in our self-normalized test statistic. }

{\normalsize The asymptotic behaviour of $\Delta $ depends on whether we are
under the null or under the alternative hypothesis. Under the null, it
follows that $\theta _{0}=0=E[f_{t}(\theta _{0})]$ and thus $\Delta =\sqrt{P}%
\hat{E}_{P}\left[ 2Y_{t}X_{t}^{\prime }\widehat{\theta }\right] .$ Under
suitable regularity conditions, we show that when (\ref{H0}) holds%
\begin{equation}
\hat{t}=\frac{\sqrt{P}\hat{E}_{P}\left[ Y_{t}X_{t}^{\prime }\widehat{\theta }%
\right] }{\sqrt{\hat{E}_{P}\left[ \left( Y_{t}X_{t}^{\prime }\widehat{\theta
}\right) ^{2}\right] }}\overset{d}{\longrightarrow }N(0,1).  \label{null}
\end{equation}%
We provide sufficient conditions for this convergence to hold in the
following result, which relies on combining our previous bounds with a
martingale CLT. We require additional
assumptions. Define $\hat{A}=\hat{E}_{P}\left[ Y_{t}^{2}X_{t}X_{t}^{\prime }%
\right] ,$ $\hat{B}=\hat{E}_{P}\left[ \sigma _{t}^{2}X_{t}X_{t}^{\prime }%
\right] ,$ and $A=B=E\left[ Y_{t}^{2}X_{t}X_{t}^{\prime }\right] ,$ where $%
\sigma _{t}^{2}=E[\left. Y_{t}^{2}\right\vert \mathcal{F}_{t-1}]$. }

{\normalsize \noindent \textbf{Assumption A9: }(i) The parameter space $%
\Theta $ is bounded and $\Pr \left( \widehat{\theta }=0\right) =0$; there
exists a $\delta >0$ such that (ii) $E\left[ \left\Vert
Y_{t}X_{t}\right\Vert ^{2+\delta }\right] <\infty $ and (iii) the following
rate conditions hold%
\begin{equation*}
\max \left\{ \left\Vert \hat{A}-A\right\Vert ,\left\Vert \hat{B}%
-B\right\Vert \right\} =o_{\mathbb{P}}(\lambda _{\min }(A))\text{ and }%
\left( \lambda _{\min }(A)\right) ^{1+\delta /2}P^{\delta /2}\rightarrow
\infty .
\end{equation*}%
In Assumption A9, $\widehat{\theta }$ is a generic estimator. Using Lasso
for $\widehat{\theta }$ might be problematic under the null, since it might
be the case that $\Pr \left( \widehat{\theta }=0\right) >0$, see, e.g., %
\citet{zhao2006model}. We argue that the Ridge estimator (\ref{Ridge}) is
more suitable for this application, as we expect deviations from the MDH to
be \textquotedblleft dense\textquotedblright\ rather than \textquotedblleft
sparse\textquotedblright\ (many small correlations, rather than large
isolated ones). }

\begin{theorem}
{\normalsize \label{Thm2}\textit{\ Let }Assumptions A1, $\lambda _{\max
}(\Sigma )\leq C<\infty ,$ \textit{and A9 hold. Under (\ref{H0}), the asymptotic
distribution in (\ref{null}) holds.} }
\end{theorem}

{\normalsize Under the alternative hypothesis, our previous expansions for $%
\beta $-mixing processes apply and under suitable conditions (see
Theorem 3.1)%
\begin{align*}
\Delta & =\sqrt{P}\left( \hat{E}_{P}\left[ 2Y_{t}X_{t}^{\prime }\hat{\theta}%
\right] -E[2Y_{t}X_{t}^{\prime }\theta _{0}]\right) +o_{\mathbb{P}}(1)
\overset{d}{\longrightarrow }N(0,\Omega _{a}),
\end{align*}%
where $\Omega _{a}=4\theta _{0}^{\prime }E[\left(
Y_{t}X_{t}-E[Y_{t}X_{t}]\right) \left( Y_{t}X_{t}-E[Y_{t}X_{t}]\right)
^{\prime }]\theta _{0}.$ Therefore, under the alternative hypothesis and
from our results%
\begin{equation*}
\hat{t}=O_{\mathbb{P}}(1)+\frac{\sqrt{P}E[2\left( X_{t}^{\prime }\theta
_{0}\right) ^{2}]}{\sqrt{\hat{E}_{P}\left[ \left( Y_{t}X_{t}^{\prime }%
\widehat{\theta }\right) ^{2}\right] }}\rightarrow \infty ,
\end{equation*}%
provided $\sqrt{P}\theta _{0}^{\prime }\Sigma \theta _{0}\rightarrow \infty
. $ Thus, the proposed out-of-sample test for the MDH rejects for $\hat{t}$
larger than the corresponding critical value from the normal distribution
(one-sided test) and it is consistent against the alternatives where $\sqrt{P%
}\theta _{0}^{\prime }\Sigma \theta _{0}\rightarrow \infty .$ This set
includes a large number of alternatives, but it does not include all
alternatives. For example, for alternatives where $E[Y_{t}X_{t}]=0$ but $%
E[\left. Y_{t}\right\vert \mathcal{F}_{t-1}]\neq 0$, our test is
inconsistent. Since $X_{t}$ is high-dimensional, the class of processes for
which $E[Y_{t}X_{t}]=0$ but $E[\left. Y_{t}\right\vert \mathcal{F}%
_{t-1}]\neq 0$ may be of less practical interest, and as a result, our test
is expected to have good power properties relative to existing tests that
only account for a handful of (linear) covariances. The next section shows
the good empirical performance of the new test. }

\subsection{Monte Carlo}

{\normalsize In order to assess the performance of our proposed test for the
MDH in Section 5.1, we run simulations under the null and the alternative,
and compare it with a testing procedure based on an OLS predictive model
(similar to the one proposed by \cite{clark2006using}) and with the Automatic
Portmanteau (AP) test proposed in \citet{escanciano2009a}. The number of
Monte Carlo simulations is 500. We implement our test with a cross-validated
Ridge estimator $\widehat{\theta }$ and a vector $X_{t}$ that includes 30
lags, their two by two interactions, their squares, cubes and fourth
exponents. We use 50\% or 80\% of the sample as the training set and 50\% or
20\% for testing (corresponding to $\pi =1$ and $\pi =0.25$ respectively).
To analyze the size of the tests, we fit a GARCH(1,1) model to EUR/USD data
obtaining the next DGP: $Y_{t}=\epsilon _{t}\sigma _{t},$ where
\begin{equation*}
\sigma _{t}^{2}=0.1+0.2\epsilon _{t-1}^{2}+0.7\sigma _{t-1}^{2},
\end{equation*}%
and where henceforth $\epsilon _{t}$ follows an iid standard normal
distribution. }

{\normalsize \noindent Table \ref{Table 2} shows that OLS-based tests and
Ridge-based tests have good and similar size for all sample sizes and
confidence levels. Also, the dependence on the sample splitting chosen is
moderate, and for most cases small.
\begin{equation*}
\fbox{TABLE 3 ABOUT HERE}
\end{equation*}%
\noindent To assess the power we consider the following DGPs: }

\begin{itemize}
\item {\normalsize AR(1)-GARCH(1,1): $Y_{t}=0.3Y_{t-1}+\epsilon_{t}%
\sigma_{t} $, where $\sigma_{t}^{2}=0.1+0.2\epsilon_{t-1}^{2}+0.7%
\sigma_{t-1}^{2}.$ }

\item {\normalsize First order exponential autoregressive model (EXP(1)): $%
Y_{t}=0.6Y_{t-1}e^{-0.5Y_{t-1}^{2}}+\epsilon_{t}.$ }

\item {\normalsize Nonlinear Moving Average model (NLMA): $%
Y_{t}=\epsilon_{t-1}\epsilon_{t-2}(1+\epsilon_{t}+\epsilon_{t-2}).$ }
%

\item {\normalsize A modified EXP(1) process with 4 autoregressive terms
AR(4)-EXP(1): $%
Y_{t}=10e^{-0.5Y_{t-1}^{2}}+0.58Y_{t-1}+0.1Y_{t-2}+0.06Y_{t-3}+0.02Y_{t-4}+%
\epsilon_{t}.$ }
\end{itemize}

{\normalsize \noindent The empirical rejection probabilities are reported in
Table \ref{Table 3}. For the AR(1)-GARCH(1,1) and EXP(1), the power of the
Ridge-based test is larger than the one of the OLS-based test for all sample
sizes and confidence levels. As our specified model does not contain Moving
Average (MA) regressors, neither of the prediction-based tests have good
power against the NLMA alternative. The AP test has excellent power against
the first two alternatives, beating the OLS and Ridge based tests. However,
for the rest of alternatives, the AP test has low power, consistent with
these alternatives presenting serial nonlinear dependence but zero or small
autocorrelations. Regarding the dependence of the tests performance on the
sample splitting chosen, we see that the power is lower when we choose a
smaller $\pi$. This is related with what we see in the previous simulations,
as the finite sample bias of the $\Delta$ distribution is larger for $\pi=1$%
, increasing the rejection probability of the tests. }

{\normalsize The last process has near zero autocovariances but it
has structure in the mean in the sense that $Y_{t}$ is correlated with
nonlinear functions of its lags. This helps to illustrate two key aspects
about the comparison between the AP and our test. First, the AP test has no
power against the alternative if the process has zero autocovariances.
Second, our test has power against the alternative if the forecast of the
process is not zero, which seems to be the case here,
see the discussion above about the consistency of our test. }

{\normalsize
\begin{equation*}
\fbox{TABLE 4 ABOUT HERE}
\end{equation*}%
}

\subsection{Empirical application to exchange rates}

{\normalsize We apply our predictive testing procedure to two sets of
exchange rate data. There is, of course, an extensive empirical literature
on the use of machine learning methods for time series predictability,
including predictability of exchange rates, see, e.g., %
\citet{plakandaras2015forecasting} and references therein. We contribute to
this extensive empirical literature with a new out-of-sample predictive
tests having justified theoretical guarantees. }

{\normalsize As the machine learner, we use cross-validated Ridge with a
vector $X_{t}$ including up to 30 lags, their interactions, squares, cubes,
and fourth powers, as in the simulations, aiming to have power against a
large set of price patterns that may have predictive power. The sample is
split into training and testing for $\pi \in \{0.25,1\}$. As we are in a
time series setup, the standard cross-validation cannot be used to choose
the regularization parameter. Among the different options, we will use
blocked cross-validation. This consists in first splitting the sample into $%
k $ blocks and then splitting each of them in training and testing. Finally,
the mean squared errors of the testing subsamples are averaged. In this
case, we split the training sample into 2 blocks ($k=2$). }

{\normalsize We compare the results of our prediction-based MDH tests with
the Automatic Portmanteau test developed in \citet{escanciano2009a}, which
has been shown to have a good performance in terms of power. As a second
benchmark, we compare with the model estimated using OLS, as in %
\citet{clark2006using}. There exists a large classical literature on testing
the MDH, see, e.g., \citet{escanciano2009a}, \citet{escanciano2009b},
including tests that detect nonlinear alternatives, see, e.g., %
\citet{escanciano2006a}. The goal of this
application is to show if the use of machine learning predictive methods
improves upon these classical methods in detecting (possibly nonlinear)
dependence in exchange rate data. The previous simulations suggest that our
method has the potential to do so even in cases where the process has no
serial linear dependence, given that features included in $X_{t}$ may
present some correlation with $Y_{t}$ (higher order serial dependence). }

{\normalsize The first dataset contains daily increments of bilateral
exchange rates obtained as in \citet{escanciano2009a} from \href{http://www.federalreserve.gov/Releases/h10/hist}%
{http://www.federalreserve.gov/Releases/h10/hist} going from 16/11/1987 to
16/11/2007 ($T=6000$ observations). We report the empirical rejections for
this data in Table \ref{Table_combined}. Taking as reference a 5$%
\%$ confidence level, none of the tests reject the MDH for the AUD, CAD,
DKK, GBP, HKD and JPY . For the HKD, the AP does not reject the null while
the OLS and Ridge tests do at $\pi =0.25$ and $\pi =1,$ respectively. This
suggests the existence of nonlinear dependence in the price series. }

{\normalsize
\begin{equation*}
\fbox{TABLE 5 ABOUT HERE}
\end{equation*}%
}

{\normalsize The second dataset consists of high-frequency data (1 minute)
of bilateral exchange rates between 2023-05-15 and 2023-07-02 ($T=6000$
observations), obtained from the Dukascopy database. For the USD/CAD,
USD/JPY and EUR/USD, none of the tests reject the null. Regarding the
GBP/CAD, all tests reject the null at a 5\% confidence level. Interestingly,
for the GBP/USD with $\pi =1$ the Ridge-based test rejects the null at 5\%,
while the AP and OLS tests do not, suggesting the existence of nonlinear
dependence detected thanks to the use of machine learning methods. For the
GBP/AUD with $\pi =1$, the AP rejects the null at 8\% while OLS and Ridge
prediction-based tests do not. The last two observations are consistent with
our simulation results as prediction-based tests showed lower power than the
AP for linear DGPs, and the power was generally larger for Ridge with $\pi
=1 $. }

{\normalsize Several conclusions emerge from this application. First,
inference may be sensitive to sample splitting when using machine learning
methods. This can be related to the size of the estimation risk, as explained
by our theoretical and simulation results throughout the
paper. We documented that for accurate confidence intervals on zero-mean
score loss functions, a small value of the out-of-sample to in-sample ratio
should be considered (e.g. $\pi =0.25)$ while when using our MDH test,
larger values of this ratio (e.g. $\pi =1$) increase the power without
significantly affecting the size. We maintain these recommendations on the
sample splitting along with a warning about the effect of structural breaks
in our methods. Second, the use of machine learning techniques, such as
Ridge, seems to bring new information regarding the predictability of some
of the exchange rates considered when using large models under the
alternative. This suggests the existence of nonlinear dependence in some of
the price series analyzed, in contrast with the lack of serial linear
dependence that traditional correlation-based tests suggest for all the
exchange rates considered except the GBP/CAD at 1-minute frequency. Given
these results, we advocate for the complementary use of the AP test and our
MDH test with large models estimated by Ridge using a not very small sample splitting ratio (e.g. $\pi =1$). %
}

\section{Conclusions}

{\normalsize The paper shows that to obtain standard asymptotic theory in
out-of-sample analysis with machine learning methods two key properties must
be satisfied simultaneously: (i) a zero-mean condition for the score of the
prediction loss function; and (ii) a fast rate on the machine learner. We
have documented theoretically and by simulations these points. We have also
found sufficient primitive conditions for $\beta $-mixing and $\alpha $-mixing time series
building on recent work by \citet{wong2020lasso} and \citet{kengne2024deep}. We have illustrated
the applicability of our results with three different applications using different loss functions and machine learners. }

{\normalsize There are several interesting venues for further research
related to our results. First, the investigation of cases where the score of
the loss function does not have a zero-mean. For example, for the absolute
mean error loss function, the score does not have a zero-mean with an asymmetric error distribution in general. We
will investigate ways to transform the loss function so that standard
inference applies. A second interesting question is the (adaptive) choice of
the sample splitting. At the moment, we can only offer a simple
recommendation based on our theoretical and simulation results. For accurate
confidence intervals with machine learning, a small value of $\pi $ such as $%
\pi =0.25$ is recommended, while for MDH testing a larger value such as $\pi
=1$ is generally preferred. These two problems are beyond the scope of this
paper, and will be investigated in future research. }

{\normalsize %
%
%
%
\bibliographystyle{chicago}
\bibliography{RPA_JBESRevApril2025}
}

{\normalsize 
}

\newpage

\section*{Figures and Tables}

{\normalsize
\begin{figure}[H]
{\normalsize \centering
\includegraphics[width=0.95\textwidth, height=5.5cm]{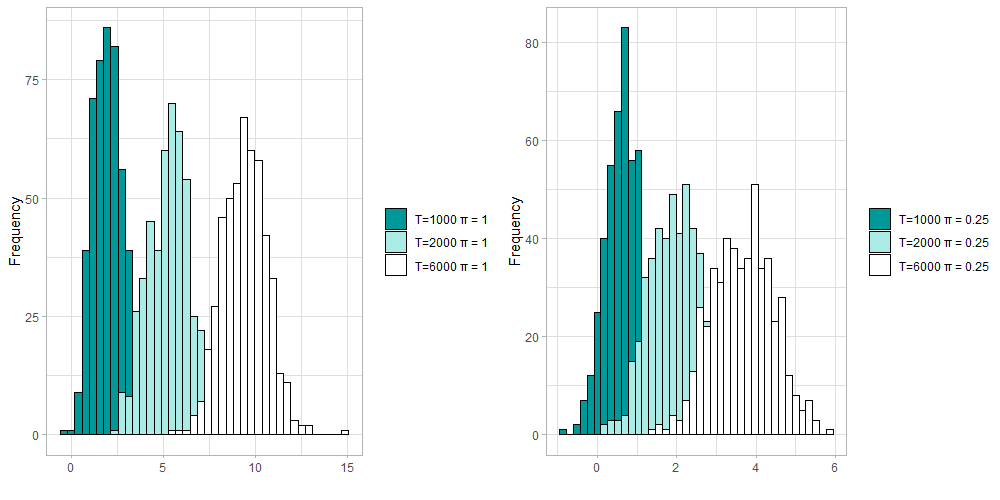}
}
\caption{ER. Sparsity Index=$s=\lceil \protect\sqrt{p}-27\rceil $, $p=T$.
Decreasing sparsity.}
\label{Figure1}
\end{figure}
}

{\normalsize
\begin{figure}[H]
{\normalsize \centering
\includegraphics[width=0.95\textwidth, height=5.5cm]{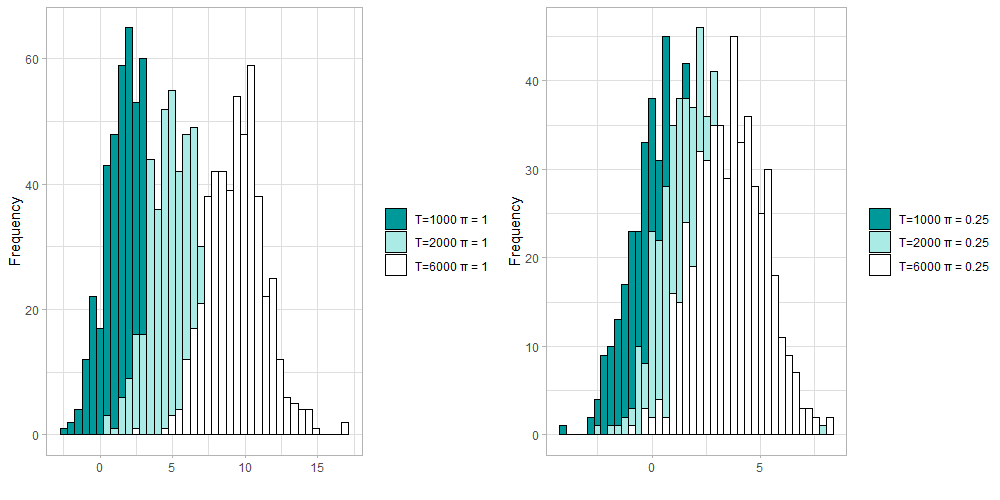}
}
\caption{$\Delta $. Sparsity Index=$s=\lceil \protect\sqrt{p}-27\rceil $, $%
p=T$. Decreasing sparsity. }
\label{Figure2}
\end{figure}
}

{\normalsize
\begin{figure}[H]
{\normalsize \centering
\includegraphics[width=0.95\textwidth, height=5.5cm]{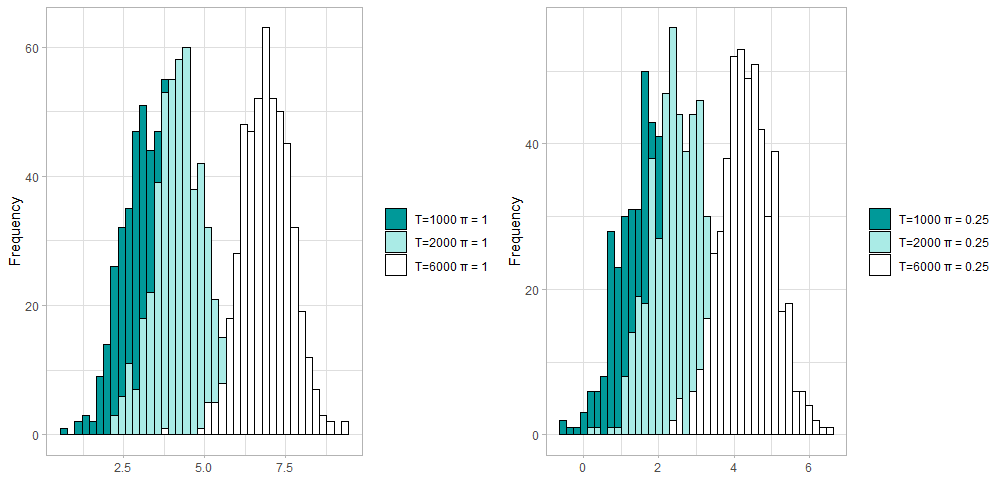}
}
\caption{$ER$. Sparsity Index=15, $p=T$. Multicollinearity.}
\label{Figure3}
\end{figure}
}

{\normalsize
\begin{figure}[H]
{\normalsize \centering
\includegraphics[width=0.95\textwidth, height=5.5cm]{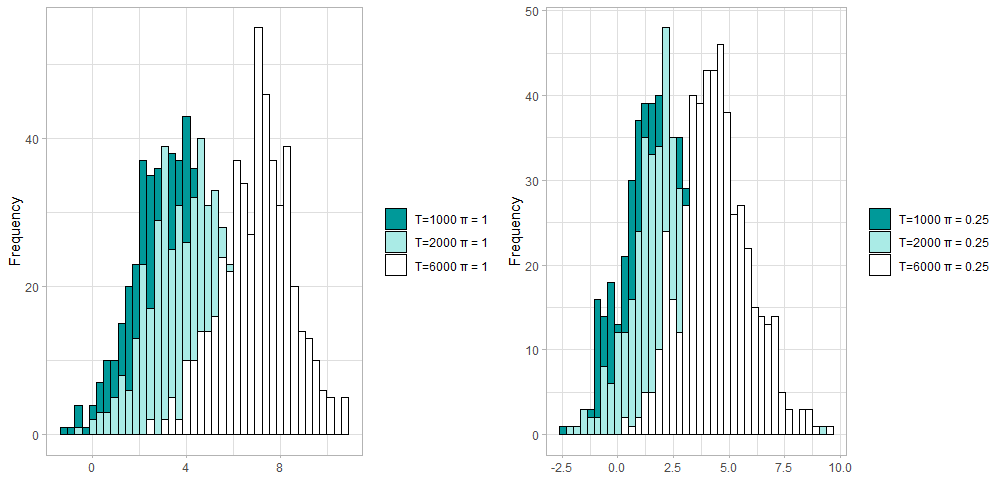}
}
\caption{$\Delta $. Sparsity Index=15, $p=T$. Multicollinearity.}
\label{Figure4}
\end{figure}
}

{\normalsize
\begin{figure}[H]
{\normalsize \centering
\includegraphics[width=0.95\textwidth, height=5.5cm]{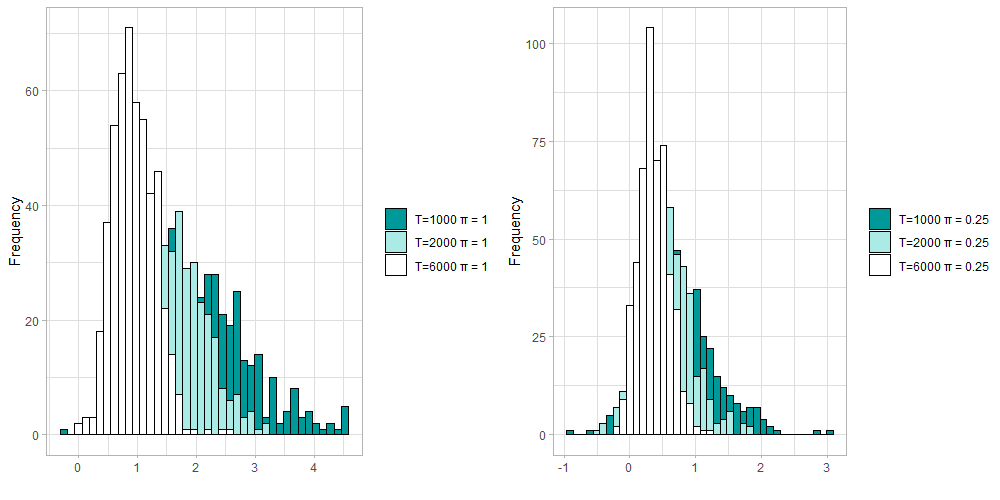}
}
\caption{ER. Sparsity Index=5, $p=T$. Fast rates. }
\label{Figure5}
\end{figure}
}

{\normalsize
\begin{figure}[H]
{\normalsize \centering
\includegraphics[width=0.95\textwidth, height=5.5cm]{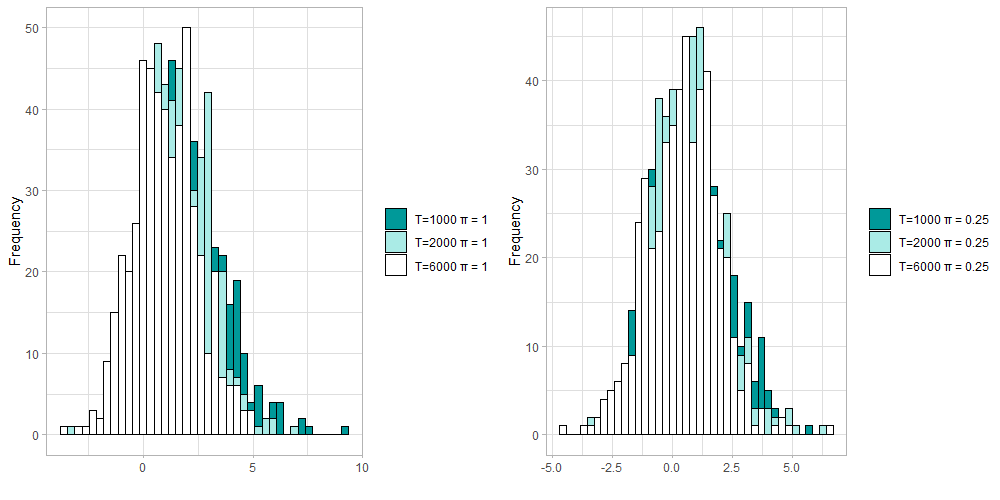}
}
\caption{$\Delta $. Sparsity Index=5, $p=T$. Fast rates. }
\label{Figure6}
\end{figure}
}

\bigskip

\begin{table}[H]
    \centering
    \resizebox{\textwidth}{!}{%
    \begin{tabular}{l c c c}
        \toprule
        \textbf{Loss} & $\mathbf{\ell(y,m)}$ & $\psi (\varepsilon_t)$ & \textbf{Zero-Mean Score} \\
        \midrule
        \textbf{MSPE} & $0.5(y - m)^2$ & $\varepsilon_t$ & Yes \\
        \textbf{MAD} & $\left| y - m \right|$ & $\mathbf{1}(\varepsilon_t \leq 0) - 0.5$ & Yes \\
        \textbf{Huber} &
        $\begin{cases}
            0.5(y - m)^2, & \text{if } |y - m| \leq \delta \\
            \delta |y - m| - 0.5\delta, & \text{if } |y - m| > \delta
        \end{cases}$ &
        $\begin{cases}
            \varepsilon_t, & \text{if } |\varepsilon_t| \leq \delta \\
            \delta \text{sign}(\varepsilon_t), & \text{if } |\varepsilon_t| > \delta
        \end{cases}$ & Yes \\
        \textbf{ASMSPE} &
        $\begin{cases}
            \alpha (y - m)^2, & \text{if } y - m \geq 0 \\
            \beta (y - m)^2, & \text{if } y - m < 0
        \end{cases}$ &
        $\begin{cases}
            \alpha \varepsilon_t, & \text{if } \varepsilon_t \geq 0 \\
            \beta \varepsilon_t, & \text{if } \varepsilon_t < 0
        \end{cases}$ & Maybe \\
        \textbf{Log-Cosh} & $\log (\cosh (y - m))$ & $\tanh(\varepsilon_t)$ & Yes \\
        \bottomrule
    \end{tabular}
    }
    \caption{Loss functions.}
    \label{Table0}
\end{table}

{\normalsize
\begin{table}[H]
{\normalsize \centering%
\resizebox{\textwidth}{!}{\begin{tabular}{rrrrrrrrrr}
\hline
\multicolumn{1}{c|}{\textbf{Sample Size}} & \multicolumn{3}{c|}{\textbf{1000}
} & \multicolumn{3}{c|}{\textbf{2000}} & \multicolumn{3}{c}{\textbf{6000}}
\\ \hline
\multicolumn{1}{c|}{\textbf{Confidence Level}} & \multicolumn{1}{c}{\textbf{0.1}} & \multicolumn{1}{c}{\textbf{0.05}} & \multicolumn{1}{c|}{\textbf{0.01}
} & \multicolumn{1}{c}{\textbf{0.1}} & \multicolumn{1}{c}{\textbf{0.05}} &
\multicolumn{1}{c|}{\textbf{0.01}} & \multicolumn{1}{c}{\textbf{0.1}} &
\multicolumn{1}{c}{\textbf{0.05}} & \multicolumn{1}{c}{\textbf{0.01}} \\
\hline
\multicolumn{1}{c|}{\textbf{Decreasing Sparsity}} &  &  &
\multicolumn{1}{r|}{} &  &  & \multicolumn{1}{r|}{} &  &  &  \\
\multicolumn{1}{c|}{$\pi =1$} & \multicolumn{1}{c}{0.628} &
\multicolumn{1}{c}{0.754} & \multicolumn{1}{c|}{0.912} & \multicolumn{1}{c}{
0.056} & \multicolumn{1}{c}{0.086} & \multicolumn{1}{c|}{0.242} &
\multicolumn{1}{c}{0.000} & \multicolumn{1}{c}{0.002} & \multicolumn{1}{c}{
0.002} \\
\multicolumn{1}{c|}{$\pi =0.25$} & \multicolumn{1}{c}{0.846} &
\multicolumn{1}{c}{0.930} & \multicolumn{1}{c|}{0.992} & \multicolumn{1}{c}{
0.642} & \multicolumn{1}{c}{0.784} & \multicolumn{1}{c|}{0.922} &
\multicolumn{1}{c}{0.258} & \multicolumn{1}{c}{0.374} & \multicolumn{1}{c}{
0.610} \\
\multicolumn{1}{c|}{\textbf{Multicollinearity}} &  &  & \multicolumn{1}{c|}{}
&  &  & \multicolumn{1}{c|}{} &  &  &  \\
\multicolumn{1}{c|}{$\pi =1$} & \multicolumn{1}{c}{0.326} &
\multicolumn{1}{c}{0.466} & \multicolumn{1}{c|}{0.736} & \multicolumn{1}{c}{
0.154} & \multicolumn{1}{c}{0.246} & \multicolumn{1}{c|}{0.458} &
\multicolumn{1}{c}{0.002} & \multicolumn{1}{c}{0.004} & \multicolumn{1}{c}{
0.024} \\
\multicolumn{1}{c|}{$\pi =0.25$} & \multicolumn{1}{c}{0.700} &
\multicolumn{1}{c}{0.824} & \multicolumn{1}{c|}{0.950} & \multicolumn{1}{c}{
0.534} & \multicolumn{1}{c}{0.642} & \multicolumn{1}{c|}{0.854} &
\multicolumn{1}{c}{0.120} & \multicolumn{1}{c}{0.168} & \multicolumn{1}{c}{
0.434} \\
\multicolumn{1}{c|}{\textbf{Fast Rates}} &  &  & \multicolumn{1}{c|}{} &  &
& \multicolumn{1}{c|}{} &  &  &  \\
\multicolumn{1}{c|}{$\pi =1$} & \multicolumn{1}{c}{0.664} &
\multicolumn{1}{c}{0.758} & \multicolumn{1}{c|}{0.914} & \multicolumn{1}{c}{
0.716} & \multicolumn{1}{c}{0.836} & \multicolumn{1}{c|}{0.956} &
\multicolumn{1}{c}{0.824} & \multicolumn{1}{c}{0.896} & \multicolumn{1}{c}{
0.968} \\
\multicolumn{1}{c|}{$\pi =0.25$} & \multicolumn{1}{c}{0.874} &
\multicolumn{1}{c}{0.936} & \multicolumn{1}{c|}{0.986} & \multicolumn{1}{c}{
0.892} & \multicolumn{1}{c}{0.944} & \multicolumn{1}{c|}{0.980} &
\multicolumn{1}{c}{0.904} & \multicolumn{1}{c}{0.952} & \multicolumn{1}{c}{
0.984} \\ \hline
\end{tabular}}  }
\caption{Confidence Intervals Coverage}
\label{Table 1}
\end{table}
}

{\normalsize
\begin{table}[H]
{\normalsize \centering%
\begin{tabular}{cccccccccc}
\hline
\multicolumn{1}{c|}{\textbf{Sample Size}} & \multicolumn{3}{c|}{\textbf{1000}
} & \multicolumn{3}{c|}{\textbf{2000}} & \multicolumn{3}{c}{\textbf{6000}}
\\ \hline
\multicolumn{1}{c|}{\textbf{Nominal Level}} & \textbf{0.1} & \textbf{0.05} &
\multicolumn{1}{c|}{\textbf{0.01}} & \textbf{0.1} & \textbf{0.05} &
\multicolumn{1}{c|}{\textbf{0.01}} & \textbf{0.1} & \textbf{0.05} & \textbf{%
0.01} \\ \hline
\multicolumn{1}{c|}{\textbf{GARCH(1,1)}} &  &  & \multicolumn{1}{c|}{} &  &
& \multicolumn{1}{c|}{} &  &  &  \\
\multicolumn{1}{c|}{OLS($\pi $=1)} & 0.098 & 0.052 & \multicolumn{1}{c|}{
0.012} & 0.082 & 0.040 & \multicolumn{1}{c|}{0.008} & 0.088 & 0.052 & 0.010
\\
\multicolumn{1}{c|}{OLS($\pi $=0.25)} & 0.132 & 0.062 & \multicolumn{1}{c|}{
0.002} & 0.100 & 0.044 & \multicolumn{1}{c|}{0.006} & 0.094 & 0.050 & 0.010
\\
\multicolumn{1}{c|}{Ridge($\pi $=1)} & 0.100 & 0.044 & \multicolumn{1}{c|}{
0.010} & 0.106 & 0.056 & \multicolumn{1}{c|}{0.010} & 0.110 & 0.046 & 0.012
\\
\multicolumn{1}{c|}{Ridge($\pi $=0.25)} & 0.102 & 0.050 &
\multicolumn{1}{c|}{0.014} & 0.124 & 0.062 & \multicolumn{1}{c|}{0.012} &
0.104 & 0.052 & 0.014 \\
\multicolumn{1}{c|}{AP($\pi $=1)} & 0.118 & 0.056 & \multicolumn{1}{c|}{0.014
} & 0.108 & 0.062 & \multicolumn{1}{c|}{0.018} & 0.096 & 0.044 & 0.010 \\
\multicolumn{1}{c|}{AP($\pi $=0.25)} & 0.156 & 0.082 & \multicolumn{1}{c|}{
0.022} & 0.116 & 0.058 & \multicolumn{1}{c|}{0.020} & 0.092 & 0.054 & 0.018
\\ \hline
\end{tabular}
}
\caption{Test performance. Size}
\label{Table 2}
\end{table}
}

{\normalsize
\begin{table}[H]
{\normalsize \centering
\resizebox{\textwidth}{!}{\begin{tabular}{cccccccccc}
\hline
\multicolumn{1}{c|}{\textbf{Sample Size}} & \multicolumn{3}{c|}{\textbf{1000}
} & \multicolumn{3}{c|}{\textbf{2000}} & \multicolumn{3}{c}{\textbf{6000}}
\\ \hline
\multicolumn{1}{c|}{\textbf{Nominal Level}} & \textbf{0.1} & \textbf{0.05} &
\multicolumn{1}{c|}{\textbf{0.01}} & \textbf{0.1} & \textbf{0.05} &
\multicolumn{1}{c|}{\textbf{0.01}} & \textbf{0.1} & \textbf{0.05} & \textbf{0.01} \\ \hline
\multicolumn{1}{c|}{\textbf{AR(1)-GARCH(1,1)}} &  &  & \multicolumn{1}{r|}{}
&  &  & \multicolumn{1}{r|}{} &  &  &  \\
\multicolumn{1}{c|}{OLS($\pi$=1)} & 0.466 & 0.346 & \multicolumn{1}{c|}{0.132
} & 0.958 & 0.932 & \multicolumn{1}{c|}{0.796} & 1.000 & 0.998 & 0.998 \\
\multicolumn{1}{c|}{OLS($\pi$=0.25)} & 0.574 & 0.416 & \multicolumn{1}{c|}{
0.160} & 0.954 & 0.878 & \multicolumn{1}{c|}{0.646} & 1.000 & 1.000 & 0.998
\\
\multicolumn{1}{c|}{Ridge($\pi$=1)} & 0.822 & 0.758 & \multicolumn{1}{c|}{
0.528} & 0.986 & 0.976 & \multicolumn{1}{c|}{0.952} & 1.000 & 1.000 & 1.000
\\
\multicolumn{1}{c|}{Ridge($\pi$=0.25)} & 0.724 & 0.604 & \multicolumn{1}{c|}{
0.326} & 0.974 & 0.942 & \multicolumn{1}{c|}{0.828} & 1.000 & 1.000 & 1.000
\\
\multicolumn{1}{c|}{AP($\pi$=1)} & 1.000 & 1.000 & \multicolumn{1}{c|}{0.994}
& 1.000 & 1.000 & \multicolumn{1}{c|}{1.000} & 1.000 & 1.000 & 1.000 \\
\multicolumn{1}{c|}{AP($\pi$=0.25)} & 0.976 & 0.934 & \multicolumn{1}{c|}{
0.786} & 1.000 & 1.000 & \multicolumn{1}{c|}{0.998} & 1.000 & 1.000 & 1.000
\\ \hline
\multicolumn{1}{c|}{\textbf{EXP(1)}} &  &  & \multicolumn{1}{c|}{} &  &  &
\multicolumn{1}{c|}{} &  &  &  \\
\multicolumn{1}{c|}{OLS($\pi$=1)} & 0.282 & 0.178 & \multicolumn{1}{c|}{0.058
} & 0.704 & 0.596 & \multicolumn{1}{c|}{0.342} & 1.000 & 1.000 & 1.000 \\
\multicolumn{1}{c|}{OLS($\pi$=0.25)} & 0.278 & 0.178 & \multicolumn{1}{c|}{
0.046} & 0.618 & 0.480 & \multicolumn{1}{c|}{0.238} & 0.998 & 0.992 & 0.976
\\
\multicolumn{1}{c|}{Ridge($\pi$=1)} & 0.568 & 0.446 & \multicolumn{1}{c|}{
0.222} & 0.912 & 0.858 & \multicolumn{1}{c|}{0.676} & 1.000 & 1.000 & 1.000
\\
\multicolumn{1}{c|}{Ridge($\pi$=0.25)} & 0.454 & 0.332 & \multicolumn{1}{c|}{
0.130} & 0.794 & 0.708 & \multicolumn{1}{c|}{0.448} & 0.998 & 0.998 & 0.994
\\
\multicolumn{1}{c|}{AP($\pi$=1)} & 0.996 & 0.988 & \multicolumn{1}{c|}{0.976}
& 1.000 & 1.000 & \multicolumn{1}{c|}{1.000} & 1.000 & 1.000 & 1.000 \\
\multicolumn{1}{c|}{AP($\pi$=0.25)} & 0.888 & 0.804 & \multicolumn{1}{c|}{
0.604} & 0.998 & 0.994 & \multicolumn{1}{c|}{0.950} & 1.000 & 1.000 & 1.000
\\ \hline
\multicolumn{1}{c|}{\textbf{NLMA}} &  &  & \multicolumn{1}{c|}{} &  &  &
\multicolumn{1}{c|}{} &  &  &  \\
\multicolumn{1}{c|}{OLS($\pi$=1)} & 0.072 & 0.036 & \multicolumn{1}{c|}{0.004
} & 0.122 & 0.080 & \multicolumn{1}{c|}{0.026} & 0.416 & 0.316 & 0.114 \\
\multicolumn{1}{c|}{OLS($\pi$=0.25)} & 0.078 & 0.030 & \multicolumn{1}{c|}{
0.004} & 0.118 & 0.052 & \multicolumn{1}{c|}{0.016} & 0.302 & 0.204 & 0.052
\\
\multicolumn{1}{c|}{Ridge($\pi$=1)} & 0.082 & 0.030 & \multicolumn{1}{c|}{
0.002} & 0.090 & 0.038 & \multicolumn{1}{c|}{0.012} & 0.318 & 0.206 & 0.058
\\
\multicolumn{1}{c|}{Ridge($\pi$=0.25)} & 0.066 & 0.024 & \multicolumn{1}{c|}{
0.002} & 0.070 & 0.028 & \multicolumn{1}{c|}{0.002} & 0.204 & 0.124 & 0.020
\\
\multicolumn{1}{c|}{AP($\pi$=1)} & 0.148 & 0.080 & \multicolumn{1}{c|}{0.026}
& 0.182 & 0.098 & \multicolumn{1}{c|}{0.026} & 0.148 & 0.082 & 0.026 \\
\multicolumn{1}{c|}{AP($\pi$=0.25)} & 0.154 & 0.070 & \multicolumn{1}{c|}{
0.016} & 0.122 & 0.072 & \multicolumn{1}{c|}{0.020} & 0.176 & 0.100 & 0.034
\\ \hline
\multicolumn{1}{c|}{\textbf{AR(4)-EXP(1)}} &  &  & \multicolumn{1}{c|}{} &
&  & \multicolumn{1}{c|}{} &  &  &  \\
\multicolumn{1}{c|}{OLS($\pi$=1)} & 0.382 & 0.294 & \multicolumn{1}{c|}{0.148
} & 0.660 & 0.598 & \multicolumn{1}{c|}{0.486} & 0.952 & 0.936 & 0.904 \\
\multicolumn{1}{c|}{OLS($\pi$=0.25)} & 0.274 & 0.214 & \multicolumn{1}{c|}{
0.102} & 0.552 & 0.478 & \multicolumn{1}{c|}{0.354} & 0.908 & 0.878 & 0.796
\\
\multicolumn{1}{c|}{Ridge($\pi$=1)} & 0.414 & 0.326 & \multicolumn{1}{c|}{
0.212} & 0.690 & 0.634 & \multicolumn{1}{c|}{0.560} & 0.994 & 0.994 & 0.990
\\
\multicolumn{1}{c|}{Ridge($\pi$=0.25)} & 0.322 & 0.250 & \multicolumn{1}{c|}{
0.128} & 0.698 & 0.614 & \multicolumn{1}{c|}{0.458} & 0.986 & 0.984 & 0.946
\\
\multicolumn{1}{c|}{AP($\pi$=1)} & 0.012 & 0.002 & \multicolumn{1}{c|}{0.002}
& 0.004 & 0.000 & \multicolumn{1}{c|}{0.000} & 0.006 & 0.004 & 0.004 \\
\multicolumn{1}{c|}{AP($\pi$=0.25)} & 0.016 & 0.012 & \multicolumn{1}{c|}{
0.012} & 0.010 & 0.004 & \multicolumn{1}{c|}{0.002} & 0.004 & 0.002 & 0.002
\\ \hline
\end{tabular}}  }
\caption{Test Performance. Power}
\label{Table 3}
\end{table}
}

\begin{table}[H]
    \centering
    \resizebox{\textwidth}{!}{
        \begin{tabular}{c|cccccc}
            \hline
            \textbf{\hspace{6pt} Daily data p-values \hspace{6pt}} & AUD/USD & CAD/USD & DKK/USD & GBP/USD & HKD/USD & JPY/USD \\
            \hline
            OLS ($\pi=1$) & 0.122 & 0.072 & 0.631 & 0.213 & 0.570 & 0.898 \\
            Ridge ($\pi=1$) & 0.339 & 0.151 & 0.784 & 0.193 & 0.041 & 0.962 \\
            Auto Portm. ($\pi=1$) & 0.330 & 0.701 & 0.360 & 0.112 & 0.300 & 0.532 \\
            \hline
            OLS ($\pi=0.25$) & 0.334 & 0.435 & 0.252 & 0.137 & 0.033 & 0.645 \\
            Ridge ($\pi=0.25$) & 0.491 & 0.356 & 0.721 & 0.154 & 0.445 & 0.698 \\
            Auto Portm. ($\pi=0.25$) & 0.257 & 0.554 & 0.109 & 0.460 & 0.186 & 0.517 \\
            \hline
        \end{tabular}
    }

    \resizebox{\textwidth}{!}{
        \begin{tabular}{c|cccccc}
            \hline
            \textbf{1 minute data p-values} & GBP/USD & GBP/CAD & GBP/AUD & USD/CAD & USD/JPY & EUR/USD \\
            \hline
            OLS ($\pi=1$) & 0.261 & 0.014 & 0.731 & 0.689 & 0.743 & 0.946 \\
            Ridge ($\pi=1$) & 0.021 & 0.011 & 0.782 & 0.924 & 0.942 & 0.530 \\
            Auto Portm. ($\pi=1$) & 0.648 & 0.022 & 0.070 & 0.649 & 0.911 & 0.368 \\
            \hline
            OLS ($\pi=0.25$) & 0.329 & 0.013 & 0.962 & 0.762 & 0.880 & 0.367 \\
            Ridge ($\pi=0.25$) & 0.189 & 0.006 & 0.958 & 0.872 & 0.979 & 0.716 \\
            Auto Portm. ($\pi=0.25$) & 0.410 & 0.048 & 0.295 & 0.636 & 0.606 & 0.294 \\
            \hline
        \end{tabular}
    }
    \caption{P-values exchange rate predictability}
    \label{Table_combined}
\end{table}

\end{document}